\documentclass[
reprint,
amssymb, 
nobibnotes, 
aps, 
pre
]{revtex4-2}
\usepackage{graphicx}	
\usepackage[caption=false]{subfig}	
\usepackage{hyperref}	
\usepackage{xcolor}	
\usepackage{bm}		
\usepackage{amsmath}	
\usepackage{wasysym}	
\usepackage{dsfont}		
\usepackage{ulem}	
\usepackage{cancel}
\usepackage{tikz}

\newcommand{\eq}{Eq.~}
\newcommand{\fig}{Fig.~}
\newcommand{\tab}{Tab.~}
\newcommand{\se}{Sec.~}
\newcommand{\app}{Appendix~}
\newcommand{\reference}{Ref.~}

\newcommand{\lr}[1]{ \left( #1 \right) }
\newcommand{\LR}[1]{ \left[ #1 \right] }
\newcommand{\llrr}[1]{ \left\{ #1 \right\} }
\newcommand{\av}[1]{ \left< #1 \right> }
\newcommand{\pav}[1]{\mkern 1.5mu\overline{\mkern-1.5mu#1\mkern-1.5mu}\mkern 1.5mu}
\newcommand{\amp}[1]{ A\left[ #1 \right] }
\newcommand{\ceil}[1]{ \left \lceil #1 \right \rceil }
\newcommand{\floor}[1]{ \left \lfloor #1 \right \rfloor }
\newcommand{\abs}[1]{ \left| #1 \right| }
\renewcommand{\vec}[1]{ \mathbf{#1} }
\newcommand{\id}{\mathds{1}}

\newcommand{\arccot}{{\mathrm{arccot}}}
\newcommand{\B}{{\mathrm{B}}}

\newcommand{\crit}{{\mathrm{crit}}}

\newcommand{\dr}{{\mathrm{dr}}}
\newcommand{\fit}{{\mathrm{fit}}}
\newcommand{\free}{{\mathrm{free}}}
\newcommand{\intra}{{\mathrm{intra}}}

\newcommand{\mSS}{{\mathrm{SS}}}
\newcommand{\NN}{{\mathrm{NN}}}
\newcommand{\pin}{{\mathrm{pin}}}
\newcommand{\rlx}{{\mathrm{relax}}}
\newcommand{\run}{{\mathrm{run}}}
\newcommand{\s}{{\mathrm{s}}}

\newcommand{\stc}{{\mathrm{stc}}}
\newcommand{\sub}{{\mathrm{sub}}}
\newcommand{\swap}{{\mathrm{swap}}}
\newcommand{\thr}{{\mathrm{thr}}}

\newcommand{\tot}{{\mathrm{tot}}}
\newcommand{\wall}{{\mathrm{wall}}}
\newcommand{\Y}{{\mathrm{Y}}}

\newcommand{\ycrit}{214}
\newcommand{\ycritB}{260}
\newcommand{\ycritZero}{(221 \pm 2)}
\newcommand{\ycritZeroFit}{(262 \pm 4)}

\definecolor{newgreen}{rgb}{0.0, 0.5, 0.0}

\begin{document}
\title{Depinning dynamics of confined colloidal dispersions under oscillatory shear}
\author{Marcel H\"ulsberg}
\email{m.huelsberg@tu-berlin.de}
\author{Sabine H. L. Klapp}
\email{sabine.klapp@tu-berlin.de}
\affiliation{Institut f\"ur Theoretische Physik, Hardenbergstr. 36, Technische Universität Berlin, D-10623 Berlin, Germany}
\date{\today}
\begin{abstract}
	Strongly confined colloidal dispersions under shear can exhibit a variety of dynamical phenomena, including depinning transitions and complex structural changes.
	Here, we investigate the behaviour of such systems under pure oscillatory shearing with shear rate $\dot{\gamma}(t) = \dot{\gamma}_0 \cos(\omega t)$, as it is a common scenario in rheological experiments.
	The colloids' depinning behaviour is assessed from a particle level based on trajectories, obtained from overdamped Brownian Dynamics simulations. 
	The numerical approach is complemented by an analytic one based on an effective single-particle model in the limits of weak and strong driving.
	Investigating a broad spectrum of shear rate amplitudes $\dot{\gamma}_0$ and frequencies $\omega$, we observe complete pinning as well as temporary depinning behaviour.
	We discover that temporary depinning occurs for shear rate amplitudes above a frequency-dependent critical amplitude $\dot{\gamma}_0^\crit(\omega)$, for which we attain an approximate functional expression.
	For a range of frequencies, approaching $\dot{\gamma}_0^\crit(\omega)$ is accompanied by a strongly increasing settling time.
	Above $\dot{\gamma}_0^\crit(\omega)$, we further observe a variety of dynamical structures, whose stability exhibits an intriguing ($\dot{\gamma}_0, \omega$) dependence.
	This might enable new perspectives for potential control schemes.
\end{abstract}
\maketitle
%
\section{Introduction}
Colloidal dispersions play a pivotal role as model systems in classical statistical mechanics \cite{lowen2013introduction}.
They are furthermore ubiquitous in everyday life and play a crucial role in numerous applications such as surface coatings, lubricants or pharmaceutical products, to name a few.
Whereas early research mainly focused on equilibrium structures, phase behaviour and the impact of static perturbations (such as confinement \cite{fortini2006phase, grandner2008freezing, rice2009structure, wilms2012motion, scarratt2021forces}), more recent studies are typically concerned with non-equilibrium effects such as dynamical phase transitions, spatio-temporal structures and non-linear macroscopic properties.
This occurs, e.g., by application of shear flow \cite{brader2010nonlinearrheology, besseling2012oscillatory, wang2016large}, or time-dependent electric \cite{zhou2013computer} or magnetic fields \cite{dillmann2013two}.
The presented study is concerned with a colloidal system subject to both, spatial confinement and oscillatory shear.

Earlier studies of (strongly) confined colloidal dispersions exposed to shear flow have focused mainly on static shearing \cite{valdez1997rheology, aerov2014driven}.
In particular, in monodisperse systems of dense colloidal layers, shear leads to effects like depinning in combination with structural changes \cite{messina2006confined, vezirov2013nonequilibrium, gerloff2016depinning}, non-monotonic constitutive flow curves \cite{gerloff2016depinning} and non-trivial thermodynamic signatures \cite{gerloff2018stochastic}.
Additional effects occur in more complex systems, e.g. laning transitions in trilayers \cite{gerloff2017shear} or density fluctuations in bidisperse systems \cite{gerloff2016depinning}.
Furthermore, extensive research has been dedicated towards driven two-dimensional (2D) colloidal systems, which can also be seen as a form of (extreme) confinement.
Prominent examples are monolayers driven over substrates by external forces \cite{hasnain2013dynamic, reichhardt2016depinning} and circular ``nanoclutches'' \cite{ortiz2018laning, gerloff2020dynamical, williams2022rheology}, i.e. ring-like configurations of a finite number of colloids, which are sheared by the application of torques.

Here we are interested in the impact of \textit{oscillatory} shearing, which, for \textit{confined} colloidal systems, has not received much attention so far \cite{cohen2004shear}.
It is, however, commonly applied in bulk systems \cite{hyun2011review, besseling2012oscillatory, fiorucci2021oscillatory}, where it is used to characterize rheological properties like storage or loss moduli \cite{brader2010nonlinearresponse}. 
Often reported phenomena in bulk systems include shear banding \cite{fiocco2013oscillatory, radhakrishnan2018shear}, shear thinning \cite{hajnal2009flow, brader2010nonlinearrheology, nicolas2016shear} and shear thickening \cite{wagner2009shear, brader2010nonlinearrheology, malkin2016shear}, to name a few.
Particular attention has been recently devoted to amorphous solids under oscillatory shear and the impact of the yielding transition \cite{parley2022meantheory, liu2022fate, priezjev2022mechanical, yeh2020glass, fielding2020elastoviscoplastic}.
While oscillatory shearing generally leads to more complex behaviour compared to static shearing, it should also allow access to system-intrinsic time scales by analyzing the system's response to different shearing frequencies.
Beyond these issues, we note that oscillatory drive in colloidal systems has also been used to study synchronization phenomena such as mode locking \cite{juniper2015microscopic}.
The present system adds a further factor due to the presence of spatial confinement.

In the present paper we are concerned with monodisperse, dense colloidal dispersions, which are confined to a narrow slit-pore and exposed to oscillatory shear acting parallel to the confining walls.
By performing overdamped Brownian Dynamics (BD) simulations in the absence of a solvent, we aim at understanding the complex collective translational and structural dynamics on the particle level. 
We regard this as a necessary first step before examining the system's rheological properties, which we leave for a future study.

We specifically restrict our research to systems consisting of only two particle layers, so called ``bilayers'', where the slit-pore width and particle density are tuned such that both layers exhibit a square-like in-plane structure in the absence of shear \cite{fortini2006phase, grandner2008freezing}.
In these systems, the application of static shear leads to three non-equilibrium steady states at small, intermediate and large shear rates \cite{vezirov2013nonequilibrium, gerloff2016depinning}.
While the first state is referred to as ``pinned'' state, where both layers are locked into each other,
the second and third state correspond to disordered and ordered ``running'' (``depinned'') states, respectively, where the two layers move in opposite directions past each other.
In addition to these differences in the translational behaviour, each state features a distinct in-plane layer structure.

Under oscillatory shear, as we show here, the system alternates periodically between the above mentioned states.
Interestingly, the question which state is included in a cycle turns out to strongly depend on the applied shear rate amplitude and frequency.
One main goal in this study is to unravel these dependencies over a large range of parameter combinations, yielding a non-equilibrium state diagram.

Another goal of this paper is to explore various system-specific (intrinsic) time scales and their interaction with the oscillation period of the externally applied shearing protocol.
One essential type of time scale in this context are ``settling times'', which denote the time span between the onset of oscillatory driving and the occurrence of stable periodic orbits of certain dynamical observables.
Interestingly, we find that these settling times can drastically increase in regions close to the dynamical depinning transition, similar to the relaxation times observed for depinning under static shear \cite{vezirov2015manipulating}.

To support the BD simulations and to interpret certain aspects of the observed collective behaviour, we establish a close connection to an effective single-particle model.
Specifically, we utilize a forced Adler-type equation \cite{adler1946study}, which we investigate analytically and numerically.
Similar (deterministic and stochastic) problems have been investigated in studies of single colloidal particles that are driven over periodic substrates in one or two dimensions \cite{risken1996fokker, pelton2004transport, reichhardt2004directional, hu2007amplitude, juniper2015microscopic, vizarim2020shapiro} with either \textit{static} [$F(t)=F$] \cite{risken1996fokker, pelton2004transport, reichhardt2004directional} or \textit{modulated} forces [$F(t)=F+F_0 \cos(\omega t)$] \cite{hu2007amplitude, juniper2015microscopic, vizarim2020shapiro}.
In our single-particle investigations, we focus on (pure) oscillatory driving [$F(t)=F_0 \cos(\omega t)$], which is a subproblem of modulated driving ($F=0$).
However, lacking full analytical solutions, previous studies \cite{hu2007amplitude, juniper2015microscopic, vizarim2020shapiro} have typically only been concerned with transport properties like average particle velocities $\left< \dot{x} \right>$, which always vanish without a constant force, as long as the substrate is symmetric and the system is noise-free.
Here we rather take a closer look at the forced oscillation's amplitudes and centers.
The latter have, in fact, been studied in several applications in the field of laser physics \cite{chow1985ring, cotteverte1994vectorial}.
There it was found that systems tend to oscillate around one of few fixed points \cite{cotteverte1994vectorial}, of which some are stable and some are unstable.
The stability of these fixed points furthermore depends not only on the underlying periodic potential but also on driving amplitude $F_0$ and frequency $\omega$.
Here we show that swapping a fixed point's stability can, in some cases, be related to the properties of the entire many-particle system, e.g., the layer structure, while transport is effectively zero.
Moreover, we demonstrate that the depinning diagram of the full many-particle system displays strong similarity with that of the single-particle system, showing that the latter captures an important part of the physics.

The rest of this paper is organized as follows. 
In \se\ref{sec:model_many_particle} we describe the model of our many-particle system and the details of our BD simulations.
In \se\ref{sec:single_particle} we motivate an effective single-particle model and present analytical solutions in limiting cases.
We then proceed by analyzing the many-particle dynamical behaviour under oscillatory shear (\se\ref{sec:temporary_depinning}), including an analysis of settling times (in \se\ref{sec:settling_time}).
Finally, \se\ref{sec:structure} is devoted to the time-dependent structural transitions within the particle layers.
We close with concluding remarks and suggestions for future research in \se\ref{sec:conclusion}.
\section{Model and simulation details} \label{sec:model}
\subsection{Sheared slit-pore many-particle system} \label{sec:model_many_particle}
\begin{figure}
	\includegraphics[width=\linewidth]
	{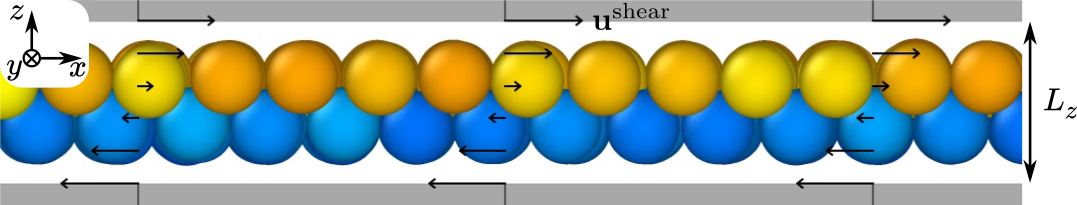}
	\caption{
		Front-view snapshot of the bilayer system confined in a narrow slit-pore environment.
		Top and bottom layer particles are colored yellow and blue, respectively.
		The confining walls (grey bars) extend infinitely in $x$- and $y$-directions.
		The linear shear velocity profile is indicated by arrows, pointing in the $x$-direction.
		Rendered with OVITO \cite{stukowski2009visualization}.
	}
	\label{fig:layout}
\end{figure}

We consider a colloidal dispersion of spherical macroions immersed in a solvent consisting of much smaller fluid molecules, as well as salt- and counterions. 
The dispersion is confined to a narrow slit-pore environment, consisting of two plane-parallel soft walls with distance $L_z$ and infinite extent in $x$- and $y$-direction (see \fig\ref{fig:layout}). 
Following previous studies \cite{klapp2007long, vezirov2013nonequilibrium, gerloff2016depinning}, we assume that the macroions interact via a combination of screened Coulomb (Yukawa) and repulsive (soft-sphere) interaction potentials, while the interaction of the macroions with the confining walls of the slit-pore is modeled via an integrated soft-sphere potential \cite{hansen2013theory}. 
We drive the system out of equilibrium by applying an oscillatory shear flow within the solvent.
In the subsequent paragraphs we first present the equation of motion, followed by a discussion of parameters and quantities of interest.
\subsubsection{Equations of motions} \label{sec:many_eq_motion}
We perform overdamped BD simulations in three dimensions, where the $i$-th macroion's position $\vec{r}_i$ ($i=1,\dots,N$) is determined by the equation of motion
\begin{equation} \label{eq:overdamped-sde}
	\dot{\vec{r}}_i = \mu \vec{F}_i + \mu \vec{\Gamma}_i + \vec{u}^{\mathrm{shear}}_i ,
\end{equation}
which we solve numerically by employing the integration algorithm from \citeauthor{ermak1975computer} \cite{ermak1975computer}. Here, $\vec{F}_i(\vec{r}_1, \dots, \vec{r}_N) = - \boldsymbol{\nabla}_i U_\tot(\vec{r}_1, \dots, \vec{r}_N)$ is the total conservative force resulting from particle-particle and particle-wall interactions acting on particle $i$. The particle-solvent coupling is contained implicitly via the mobility $\mu$, the stochastic Brownian force $\vec{\Gamma}_i(t)$ and the shear flow $\vec{u} ^{\mathrm{shear}}_i(z_i, t)$. In the following, we discuss in more detail the various terms entering \eq\eqref{eq:overdamped-sde}.

To start with, we define the total potential energy as
\begin{equation}
U_\tot = \frac{1}{2} \sum_i \sum_{j \neq i} \LR{U_\Y(r_{ij}) + U_\mSS(r_{ij})} + \sum_i U_\wall(z_i) ,
\end{equation}
where the Yukawa potential $U_\Y(r_{ij})$ and the soft-sphere potential $U_\mSS(r_{ij})$, both depending on the particle distance $r_{ij} = | \vec{r}_i - \vec{r}_j |$, are the particle-particle interaction potentials and $U_\wall(z_i)$ is the particle-wall interaction potential. The Yukawa potential arises within the framework of the Derjaguin-Landau-Verwey-Overbeek (DLVO) approximation \cite{hansen2013theory}.
It describes, on a mean-field level, the electrostatic interaction of the macroions which is screened by salt- and counterions in the solvent, yielding
\begin{equation} \label{eq:U_Yukawa}
U_\Y(r_{ij}) = \epsilon_\Y d \frac{e^{-\kappa r_{ij}}}{r_{ij}} .
\end{equation}
In \eq\eqref{eq:U_Yukawa}, $\kappa$ is the inverse Debye screening length, $\epsilon_\Y$ is the interaction strength and $d$ is the particle diameter. The soft-sphere potential accounts for the steric repulsion between the macroions and is chosen as the repulsive part of the Lennard-Jones potential,
\begin{equation} \label{eq:U_SS}
U_\mSS(r_{ij}) = 4 \epsilon_\mSS \left( \frac{d}{r_{ij}} \right)^{12} ,
\end{equation}
where $\epsilon_\mSS$ is the soft-sphere interaction strength.
Following previous studies \cite{vezirov2013nonequilibrium, gerloff2016depinning}, the particle-particle interactions are truncated for distances $r_{ij} > r_c = 2.96d$ and shifted accordingly to ensure continuous energies and forces \cite{nicolas1979equation, vezirov2013nonequilibrium}.

The particle-wall interaction is modeled via an integrated soft-sphere potential, where the wall is assumed to consist of homogeneously distributed wall particles with mean diameter $d$. Integrating \eq\eqref{eq:U_SS} over a half-space for upper and lower wall (located at $z \gtrless \pm L_z/2$), the particle-wall interaction potential reads
\begin{equation}
U_\wall(z_i) = \frac{4\pi}{5} \epsilon_\wall \left[ 
\left( \frac{d}{L_z/2 - z_i} \right)^9 + 
\left( \frac{d}{L_z/2 + z_i} \right)^9
\right] ,
\end{equation}
where $\epsilon_\wall$ is the wall interaction strength. This kind of fluid-wall interaction is widely adopted \cite{hansen2006theory, klapp2007long, grandner2008freezing}, although there are different conventions in how to choose the \mbox{prefactor}. Here, we stick with the definition from \reference\cite{grandner2008freezing}.

Within the framework of our BD simulations, the influence of the solvent on the macroions leads to friction and noise due to thermal fluctuations as well as to drag due to the shear flow. 
The former two are described by the mobility $\mu$ (or friction coefficient $\mu^{-1}$) and the stochastic Brownian force $\vec{\Gamma}_i(t)$, which has zero mean, i.e. $\av{\vec{\Gamma}_i} = 0$, and is delta-correlated, $\av{\vec{\Gamma}_i (t) \vec{\Gamma}_j(t')} = 2 \frac{k_\B T}{\mu} \delta_{ij} \delta(t-t') \id$. 
Within the integration scheme from \citeauthor{ermak1975computer} \cite{ermak1975computer}, the stochastic force leads to random, Gaussian-distributed displacements with zero mean and variance $2 D_0 \Delta t$ for each spatial coordinate after each time step. 
Here, $D_0 = \mu k_\B T$ is the short-time diffusion coefficient and $\Delta t$ is the discretized time step size. 
Furthermore, $D_0$ defines the so-called Brownian time $\tau_\B = d^2/D_0$, which we use as the reference time scale in our simulations, together with the length scale $d$ (particle diameter) and the energy scale $k_\B T$ (thermal energy).

Finally, we model the drag from the shear flow as $\vec{u}^{\mathrm{shear}}_i(z_i, t) = \dot{\gamma}(t) z_i \vec{e}_x$, describing a (spatially) linear gradient profile in $z$-direction with flow in $x$-direction. This ansatz follows previous studies \cite{messina2006confined, besseling2012oscillatory, vezirov2013nonequilibrium, gerloff2016depinning}, where hydrodynamic interactions are neglected.
This is typically justified by the conjecture that hydrodynamic interactions affect only the time scales but not the overall behaviour of these systems \cite{vezirov2013nonequilibrium, gerloff2016depinning}.

We justify the assumption of a linear shear velocity profile by the fact that the colloidal particles are located far enough from the walls.
That is, the Coulomb repulsion between the real, charged colloids and charged walls is sufficiently strong, i.e. corresponding to large $\epsilon_\wall$ in our coarse-grained model (for a more detailed discussion of the underlying charged system, see, e.g. \reference\cite{grandner2008freezing}).
Thus, it seems safe to assume that the motion of the colloids is not directly coupled to that of the wall particles.
Note, however, that close to the walls the real flow profile can be non-linear \cite{delhommelle2003effects}. 
In addition, the assumption of a linear shear profile might also break down for fast shearing, because the flow field might then no longer be able to adapt to the wall movement instantaneously.
We leave the consideration of hydrodynamic interactions for future studies, as it would go beyond the scope of this work.

In this study, we apply an oscillatory, harmonic shear protocol with shear rate
\begin{equation} \label{eq:shear_protocol}
	\dot{\gamma}(t) = \dot{\gamma}_0 \cos(\omega t) ,
\end{equation}
where $\dot{\gamma}_0$ is the shear rate amplitude and $\omega$ the frequency. 
The shear rate is chosen such that the corresponding shear strain $\gamma(t) = \frac{\dot{\gamma}_0}{\omega} \sin(\omega t)$ disappears at time $t=0$. Thus, we start the simulation from an initially unsheared system. The initial equilibrium configuration was relaxed beforehand in the absence of shear flow ($\dot{\gamma}=0$) for 100 Brownian times \cite{gerloff2016depinning}.
\subsubsection{Parameters} \label{sec:parameters}
Following previous studies \cite{gerloff2016depinning}, we consider a strongly confined colloidal dispersion with macroion density $\rho = 0.85 d^{-3}$ (volume fraction $\phi = \rho \frac{\pi}{6} d^3 \approx 0.45$) and wall distance $L_z = 2.2 d$. 
Under these conditions, the macroions form, in equilibrium, two crystalline layers with quadratic in-plane order and area density $\rho_\mathrm{area} = \frac{1}{2} \rho L_z \approx 0.94 d^{-2}$ (area fraction $\phi_\mathrm{area} = \rho_\mathrm{area} \frac{\pi}{4} d^2 \approx 0.73$) \cite{fortini2006phase}.
Considering a total of $N=1058$ macroions, we use a system box with side lengths $L_x = L_y = \sqrt{N/\rho L_z} = 23.79 d$, applying periodic boundary conditions in the $x$- and $y$-direction.

Regarding the interaction potentials, we stick to the same model parameters as in \reference\cite{gerloff2016depinning}, i.e. the interaction strengths $\epsilon_\Y = 123.4 k_\B T$, $\epsilon_\mSS = \epsilon_\wall = 1 k_\B T$ as well as the inverse Debye screening length $\kappa = 3.216 d^{-1}$.
These were chosen based on real suspensions of charged silica pellets \cite{klapp2007long}. 
The particle-particle interactions are truncated beyond a cut-off radius $r_{\mathrm{c}} = 2.96 d$ and are shifted accordingly \cite{vezirov2013nonequilibrium}.

Throughout this study, we consider shear rate amplitudes in the range $\dot{\gamma}_0 \tau_\B = 2.14 \cdot (10^1 \dots 10^3)$, spanning two orders of magnitude around the known critical depinning shear rate $\dot{\gamma}_0^\crit \tau_\B \approx \ycrit$ for this system at constant shearing \cite{gerloff2016depinning}. 
Similarly, we consider shearing frequencies in the range $\omega \tau_\B = 5.15 \cdot (10^{-1} \dots 10^4)$, spanning five orders of magnitude.
The discretized time step size is chosen smaller than $\Delta t \leq 10^{-5} \tau_\B$ for each simulation individually, depending on the shear rate amplitude and frequency (for details, see \app\ref{app:dt}).

To account for remaining finite-size effects in system-integrated observables like the center-of-mass positions of each layer (see below), we carry out simulations for 10-100 realizations, each starting from the same initial configuration. 
Further, each simulation is carried out for 10-100 oscillation periods $\tau_\omega = 2\pi/\omega$, depending on the settling time for each specific parameter combination $(\dot{\gamma}_0, \omega)$ (see discussion in \se\ref{sec:settling_time} and \app\ref{app:settling_time}).
\subsubsection{Quantities of interest} \label{sec:quantities_of_interest}
In this section we define various system-averaged observables that we use to characterize the (macroscopic) state of the system.
It is important to note that all of these quantities are functions of time due to the explicit time dependence of the driving. 
Hence, the system does not approach a (non-equilibrium) steady state, which could be described by a simple time average. 
Instead, for each simulation we compute different (noise) realizations (identical settings but different random seeds), which we refer to as \textit{ensembles}.
In later sections, we then analyze the following quantities based on their individual noise realizations as well as their ensemble average $\left< \cdots \right>$, both of which keep their time dependence.

As mentioned in the previous section, we are dealing with a system of two crystalline particle layers.
Under strong shear, these layers eventually start moving in opposite directions.
Following previous studies \cite{gerloff2016depinning}, we describe this movement by calculating center-of-mass layer positions
\begin{equation} \label{eq:Rn}
\vec{R}_m = \frac{1}{N_m} \sum_{i=1}^N H_m(z_i) \vec{r}_i ,
\end{equation}
where $m$ is the index of the bottom ($m=1$) and top ($m=2$) layer, $N_m = \sum_i^N H_m(z_i)$ is the number of particles in the $m$-th layer and $H_m(z_i)$ is the layer identification function, which we define as
\begin{equation} \label{eq:layer_id}
H_m(z_i) = \begin{cases} 
1, \quad & \mathrm{if} \quad \hat{z}_{m-1} < z_i < \hat{z}_m , \\
0, \quad & \mathrm{else} .
\end{cases}
\end{equation}
Here, $\hat{z}_m$ denotes the boundary between the $m$-th and $(m+1)$-th layer. 
In our bilayer system, the boundaries are located at $\hat{z}_0 = -L_z/2, \hat{z}_1 = 0, \hat{z}_2 = L_z/2$. 
Finally, we are interested in the relative motion of the layers
\begin{equation} \label{eq:dR}
\Delta \vec{R} = \vec{R}_2 - \vec{R}_1 .
\end{equation}

To evaluate the lateral structure within the layers, we calculate (intra-layer) angular bond order parameters.
Specifically, we define the order parameter \cite{vezirov2013nonequilibrium} of the $i$-th particle with symmetry $n$ ($=4,6$) by
\begin{subequations}
\begin{align}
\psi_{n,i} &= \frac{1}{N_i^\NN} \left\vert \sum_j^{N_i^\NN} e^{i n \theta_{ij}} \right\vert \\
           &= \frac{1}{N_i^\NN} \left\vert \sum_{j=1}^N G(z_i, z_j) \Theta(r_\NN - r_{ij}) e^{i n \theta_{ij}} \right\vert ,
\end{align}
\end{subequations}
where the sum over $j$ only counts the $N_i^\NN = \sum_{j=1}^N G(z_i, z_j) \Theta(r_\NN - r_{ij})$ nearest neighbours of particle $i$ within the same layer and $\theta_{ij}$ is the enclosing angle between the connection vector $\vec{r}_{ij}$ and the $x$-axis. 
Another particle $j$ is considered a nearest neighbour if it is located within the nearest neighbour radius $r_\NN$, which is chosen as the location of the first minimum of the (instantaneous) pair correlation function (see \app\ref{app:pair_correlation}).
To ensure only particles within the same layer are counted as nearest neighbours, we define the function $G(z_i, z_j) = \sum_{m=1}^{N_L} H_m(z_i) H_m(z_j)$, which is one if both particles $i$ and $j$ are located within the same layer and zero if not.

Finally, averaging the angular bond order parameter over a whole layer and then across layers yields
\begin{equation} \label{eq:psi_n}
\psi_n = \frac{1}{N_L} \sum_{m=1}^{N_L} \frac{1}{N_m} \sum_{i=1}^{N} H_m(z_i) \psi_{n,i} .
\end{equation}
By calculating $\psi_n$ as function of time we monitor the degree of $n$-fold symmetry of a given in-plane structure.
A perfect square or hexagonal lattice leads to $\psi_4 = 1$ or $\psi_6 = 1$, respectively.
\subsection{Driven single particle in a one-dimensional periodic potential} \label{sec:single_particle}
As we will show in \se\ref{sec:temporary_depinning}, some aspects of the many-particle dynamics, particularly the relative layer motion, can be understood in terms of a simplified model involving a single particle that is exposed to a periodic substrate potential and actuated by a time-dependent force.
In the following we present this model in detail and discuss some analytical and numerical results.
\subsubsection{Equation of motion}
Our simple model targets the dynamics of the $x$-component of the relative center-of-mass position [see \eq\eqref{eq:dR}], which allows for a one-dimensional (1D) description of the resulting ``effective'' particle.
This particle is subject to an oscillatory driving force $F_\dr(t) = F_{\dr,0} \cos{\omega t}$ with amplitude $F_{\dr,0}$ and frequency $\omega$, which mimicks the externally applied shear flow in the many-particle system. 
Furthermore, we expose the particle to a sinusoidal substrate potential
\begin{equation} \label{eq:Vsub}
	V_\sub(x) = V_{\sub,0} \left[ 1 - \cos \left( \frac{2\pi}{a} x \right) \right] 
\end{equation}
with spatial period $a$ and potential height $2V_{\sub,0}$. 
Compared to a true particle in the top (or bottom) layer of the many-particle system, this substrate potential represents the interactions of the particle with the particles of the neighbouring layer.
This analogy is expected to work well when the lateral structure is strongly pronounced. 
A similar mapping from many-particle to effective single-particle motion has been done in a previous work for a constant driving force \cite{gerloff2016depinning}.

We describe the motion of the effective single particle via the overdamped Langevin equation
\begin{equation} \label{eq:sde_1d}
	\dot{x} = - F_{\sub,0} \sin \left( \frac{2\pi}{a} x \right)
	+ F_{\dr,0} \cos(\omega t) + \Gamma(t) ,
\end{equation}
where $x$ is the (1D) position of the particle, $F_{\sub,0} = V_{\sub,0} 2\pi/a$ is the amplitude of the substrate force and $\Gamma(t)$ is a stochastic Brownian force.
In \eq\eqref{eq:sde_1d}, all quantities have been non-dimensionalized as described in \se\ref{sec:many_eq_motion}.
Note that colloidal particles driven over 1D periodic substrates [described by \eq\eqref{eq:sde_1d} with varying $F_\dr(t)$] have been studied before, e.g. with static drive $F_\dr(t)=F_\dr$ (deterministically and with noise) \cite{risken1996fokker, pelton2004transport} or modulated drive $F_\dr(t)=F_\dr + F_{\dr,0}\cos(\omega t)$ (in the deterministic \cite{hu2007amplitude} and stochastic case \cite{juniper2015microscopic}).
Here we focus on the case of oscillatory drive alone.

Mathematically, \eq\eqref{eq:sde_1d} represents a stochastic, nonlinear differential equation, which does not possess a full analytical solution for the entire parameter space of ($F_{\dr,0}, \omega$) combinations.
In the following we focus on the deterministic case ($\Gamma=0$), which has been shown to be useful already for static drives \cite{gerloff2016depinning}.
This leaves us with a forced Adler-type equation \cite{adler1946study}, which has been studied, e.g. in the context of mode-locking in ring laser gyros \cite{chow1985ring} or in lasers with one or two stable eigenstates \cite{cotteverte1994vectorial}.

\subsubsection{Limiting cases}
We first consider the case that the confinement of the particle by the potential barriers $2V_{\sub,0}$ dominates the impact of the driving force.
In this case (also studied in \reference\cite{cotteverte1994vectorial}), one can linearize the substrate force [$\sin \lr{2\pi x/a} \approx 2\pi x/a$] at the potential minimum $x_{\min} = 0$.
For an initial position $x(0)=x_0$ close to the minimum ($|x_0 - x_{\min}| \ll a/2$) this approximation holds, since the particle stays close to the minimum. Thus, we obtain the pinned solution
\begin{equation} \label{eq:xPinned}
x^\pin(t) = x_{\max}^\pin \sin(\omega t - \varphi) + \underbrace{\lr{x_0 - x_{\stc,0}} e^{-t/\tau_\sub}}_{x_\stc(t)}
\end{equation}
where
\begin{subequations} \label{eq:xPinnedDefs}
\begin{align}
x_{\max}^\pin(F_{\dr,0}, \omega)   &= \frac{a}{2\pi} \frac{F_{\dr,0}}{F_{\sub,0}} \frac{1}{\sqrt{1 + (\omega \tau_\sub)^2}} \label{eq:xmaxPinned} , \\
\varphi(\omega)     		  &= - \frac{\pi}{2} + \arctan(\omega \tau_\sub) \label{eq:phiPinned} , \\
x_{\stc,0}(F_{\dr,0}, \omega) &= \frac{F_{\dr,0} \tau_\sub}{1 + (\omega \tau_\sub)^2} \label{eq:xstcPinned} .
\end{align}
\end{subequations}
As revealed by \eq\eqref{eq:xPinned}, at long times the particle performs a sinusoidal motion with amplitude $x_{\max}$, driving frequency $\omega$ and phase shift $\varphi$. This long-time solution is complemented by a short-time correction (stc) term. 
The latter falls off exponentially with decay time
\begin{equation} \label{eq:tau_sub}
\tau_\sub = \frac{1}{2\pi} \frac{a}{F_{\sub,0}} ,
\end{equation}
which we subsequently call {\it substrate relaxation time}, as it depends solely on parameters of the substrate potential [compare \eq\eqref{eq:Vsub}]. 
Furthermore, pure exponential decay is recovered in the case of vanishing driving force $F_{\dr,0}=0$ and small initial displacements $x_0 \neq 0$, where the particle creeps down to the potential minimum.

At larger driving amplitudes $F_{\dr,0}$, the particle eventually overcomes the potential barrier, yielding (temporarily) depinned solutions.
In the case of very large $F_{\dr,0}$, the particle behaves almost like a free particle ($F_{\sub,0}=0$), which would perform a sinusoidal motion 
\begin{equation} \label{eq:xFree}
	x^\free(t) = x_{\max}^\free \sin(\omega t) + x_0
\end{equation}
with amplitude
\begin{equation} \label{eq:xmaxFree}
	x_{\max}^\free(F_{\dr,0}, \omega) = \frac{F_{\dr,0}}{\omega}
\end{equation}
around its initial position $x(0)=x_0$.
For nonzero, but small amplitudes of the substrate force $0 < F_{\sub,0} \ll F_{\dr,0}$, the particle performs the same (fast) sinusoidal motion
\begin{equation} \label{eq:xRunning}
x^\run(t) = x_{\max}^\free \sin(\omega t) + \pav{x}(t)
\end{equation}
plus a (slow) modulation of the oscillation center (or period average, denoted by the bar), which becomes constant at long times [for details, see \app\ref{app:xmean}, \eq\eqref{eq:theta_solution}].
We denote this behaviour as the ``running'' case.
Due to symmetry considerations, the particle either oscillates around the potential minimum ($\pav{x}=0$) or maximum ($\pav{x}=a/2$), depending on the combination of $F_{\dr,0}$ and $\omega$.
\subsubsection{Depinning in the full parameter space} \label{sec:depinning_single_particle}
Given the analytical solutions for limiting cases discussed in the previous paragraph, we now aim at determining the \textit{critical} driving amplitude at which the particle becomes (temporarily) depinned upon increasing $F_{\dr,0}$ from zero.
We note that in the limit $\omega \rightarrow 0$, i.e. for static drive, the critical force is well known; here one has $F_{\dr,0}^\crit(\omega \rightarrow 0) = F_{\sub,0}$ \cite{adler1946study}.
For finite $\omega$, however, the characterization of depinning is not straightforward.

In this study, we evaluate the particle's depinning behaviour based on its oscillation amplitude $x_{\max}$ in the long-time limit (including an evaluation of the settling time, see \app\ref{app:amplitude} and \ref{app:settling_time}).
At any combination of $F_{\dr,0}$ and $\omega$, we classify a long-time solution $x(t)$ as (temporarily) depinned, if the particle covers a distance larger than one spatial period of the substrate potential ($2 x_{\max} > a$).
In this case, it crosses two ore more potential wells/hills within each oscillation period.
In contrast, a pinned solution ($2 x_{\max} \leq a$) visits only one potential well/hill.
Note that the choice of the threshold $x_{\max} = a/2$ is arbitrary but, at the same time, plausible, as we will see later on.

Due to the lack of a full analytical solution for the entire $(F_{\dr,0}, \omega)$ parameter space, we solve \eq\eqref{eq:sde_1d} (with $\Gamma=0$) numerically to obtain trajectories $x(t)$, which we then use to calculate $x_{\max}$ in the long-time limit.
Note that although the long-time solutions are, in general, not sinusoidal, all of them are $\tau_\omega$-periodic and oscillate around either $\pav{x}=0$ or $\pav{x}=a/2$ (see also \fig\ref{fig:sd_xmean_single_particle} and the corresponding discussion in \app\ref{app:xmean}).

\begin{figure}
	\includegraphics[width=\linewidth]
	{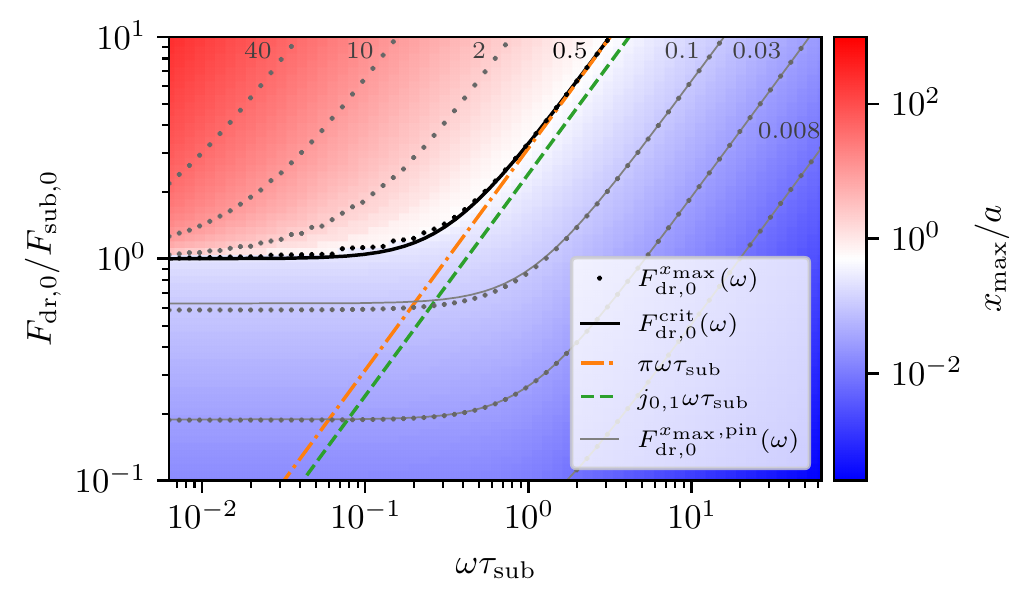}
	\caption{
		Depinning state diagram of the deterministic single-particle model in the $(F_{\dr,0}, \omega)$ parameter space based on long-time oscillation amplitudes $x_{\max}$. 
		Blue regions indicate constantly pinned solutions ($x_{\max} < a/2$), red regions (temporarily) depinned solutions ($x_{\max} > a/2$) and white regions the transition area in-between ($x_{\max} \approx a/2$).
		Contour lines $F_{\dr,0}^{x_{\max}}(\omega)$ are drawn as dots for several values of $x_{\max}/a$ (annotations).
		In the pinned region, the contour lines are well approximated by \eq\eqref{eq:contour_line_pinned} (thin solid lines).
		The contour line labeled $F_{dr,0}^\crit(\omega)$ at $x_{\max} = a/2$ marks the critical driving force amplitude [\eq\eqref{eq:Fdr0_crit}, thick solid line], which separates the pinned and depinned regions.
		At large frequencies, $F_{\dr,0}^\crit(\omega)$ becomes linear (dash-dotted orange line), at slightly larger $F_{\dr,0}$ than the first swap of the particle's oscillation center (see \app\ref{app:xmean}) from $\pav{x}=0$ to $\pav{x}=a/2$ (dashed green line).
	}
	\label{fig:sd_single_particle}
\end{figure}

In \fig\ref{fig:sd_single_particle}
we show the numerically obtained oscillation amplitudes $x_{\max}(F_{\dr,0}, \omega)$ in the form of a heatmap.
We define the critical driving amplitude $F_{\dr,0}^\crit(\omega)$, which separates the pinned (blue) and depinned (red) regions in the $(F_{\dr,0}, \omega)$ parameter space, as the contour line at $x_{\max}=a/2$ (black dots).
One clearly observes a strong dependence on the frequency for $\omega \tau_\sub \gtrsim 0.4$.
Aiming to find a mathematical expression for $F_{\dr,0}^\crit(\omega)$, we first formulate an expression for the contour lines at smaller $x_{\max}$.
In the pinned region, these contour lines (gray dots) are well approximated by
\begin{equation} \label{eq:contour_line_pinned}
	F_{\dr,0}^{x_{\max}, \pin}(\omega) = F_{\sub,0} \frac{x_{\max}}{a/2\pi} \sqrt{1 + \lr{\omega \tau_\sub}^2} ,
\end{equation}
which is obtained from \eq\eqref{eq:xmaxPinned} by solving with respect to $F_{\dr,0}$.
Equation~\eqref{eq:contour_line_pinned} works well up to $x_{\max} \lesssim 0.1a$ (thin gray lines), but starts to fail for larger $x_{\max}$.
However, it still captures the contour lines' shape (up to $x_{\max}=a/2$) quite well.
Moreover, \eq\eqref{eq:contour_line_pinned} describes the high-frequency limit $F_{\dr,0}^{x_{\max}, \pin}(\omega) \rightarrow x_{\max} \omega$ correctly, independent of $x_{\max}$.
This goes along with our observation that, at large $\omega$, the particle motion always seems to approach the running solution [\eq\eqref{eq:xRunning}].
The contour line of a free (or running) particle (for $x_{\max}=a/2$) is drawn as dash-dotted orange line in \fig\ref{fig:sd_single_particle}.

Motivated by these observations we introduce a fit parameter $\alpha$ into \eq\eqref{eq:contour_line_pinned}, specifically
\begin{equation} \label{eq:contour_line_fit}
	F_{\dr,0}^{x_{\max}}(\omega) = \alpha^{-1} F_{\sub,0} \frac{x_{\max}}{a/2\pi} \sqrt{1 + \lr{\alpha \omega \tau_\sub}^2} .
\end{equation}
We find that as $x_{\max}$ approaches the depinning threshold $a/2$, $\alpha$ changes from $1$ to $\pi$. 
This leaves us with an approximation for the critical driving amplitude at $x_{\max} = a/2$
\begin{equation} \label{eq:Fdr0_crit}
	F_{\dr,0}^\crit(\omega) = F_\sub \sqrt{1 + \left( \pi \omega \tau_\sub \right)^2} ,
\end{equation}
shown as thick black line in \fig\ref{fig:sd_single_particle}.
In the limit $\omega \rightarrow 0$, the right side of \eq\eqref{eq:Fdr0_crit} reduces to $F_{\sub,0}$, consistent with the known solution in the static case \cite{adler1946study}.
\eq\eqref{eq:Fdr0_crit} represents a very good description of our numerical results for both small and large frequencies with minor deviations at intermediate frequencies $\omega \tau_\sub \in [0.02, 0.4]$.
There, the real transition line exhibits steps, which become less pronounced at larger frequencies.

As a final note, the critical driving amplitude defined here as contour line $x_{\max}=a/2$ 
is consistent with the line of points, where the particle's oscillation center changes its stability from $\pav{x}=0$ to $\pav{x}=a/2$ for the first time (dashed green line, compare \app\ref{app:xmean}, \fig\ref{fig:sd_xmean_single_particle}).

At the end of this section, we briefly comment on the effects of (weak) noise on the depinning behaviour, i.e. $\Gamma \neq 0$ in \eq\eqref{eq:sde_1d}.
Weak noise hereby means that the thermal energy is assumed to be much smaller than the (substrate) potential barrier ($k_\B T/2V_{\sub,0} \ll 1$).
In this case, one could analogously define a critical driving amplitude $F_{\dr,0}^\crit(\omega)$ based on the contour line $\av{x_{\max}} = a/2$ ($x_{\max}$ becomes a stochastic variable, hence the average).
While the overall shape of $F_{\dr,0}^\crit(\omega)$ stays roughly the same in the presence of noise, $F_{\dr,0}^\crit(\omega \rightarrow 0)$ dips below $F_{\sub,0}$, as expected from earlier studies \cite{hasnain2013dynamic}.
Additionally, the sharp transition from pinning ($x_{\max} \leq a/2$) to depinning ($x_{\max} > a/2$) at small frequencies becomes smooth.
Apart from these effects, the overall behaviour remains the same.
\section{Simulation results} \label{sec:simulation_results}
We now turn to a discussion of our numerical results for the colloidal bilayer system described in \se\ref{sec:model_many_particle}.
\subsection{Depinning at static shearing ($\omega=0$)} \label{sec:constant_shearing}
As a reference for our subsequent discussion of oscillatory shear [$\dot{\gamma}(t)=\dot{\gamma}_0 \cos(\omega t)$], it is worth to briefly review its depinning behaviour under constant shear, which is recovered formally at $\omega=0$. 
Earlier studies \cite{vezirov2013nonequilibrium, gerloff2016depinning} discovered that the sheared system exhibits different non-equilibrium steady states in the long-time limit ($t > 50\tau_\B$), depending on the magnitude of $\dot{\gamma}_0$.
Below a critical shear rate $\dot{\gamma}_0 < \dot{\gamma}_0^\crit \approx \ycrit \tau_\B^{-1}$, both particle layers are locked into each other (pinned states).
Consequently, the center-of-mass of each layer is at rest, and thus, $\av{\Delta R_x}=const$ [for definition of $\Delta R_x$, see \eq\eqref{eq:dR}].
On the contrary, above $\dot{\gamma}_0 \geq \dot{\gamma}_0^\crit$, the layers slide past each other (depinned or running states) and $\av{\Delta R_x}$ increases with time.
This depinning transition is accompanied by a change of the lateral structure of both layers.
In the pinned state, particles arrange in a crystalline structure with square in-plane order.
For depinned states, in contrast, this order fades off resulting in a (fluid-like) disordered structure, followed by a recrystallization with hexagonal order beyond a second critical shear rate $\dot{\gamma}_0 \geq \dot{\gamma}_0^{\crit,2} \approx \ycritB \tau_\B^{-1}$.
\subsection{Temporary depinning at oscillatory shearing} \label{sec:temporary_depinning}
\begin{figure}
	\includegraphics[width=\linewidth]
	{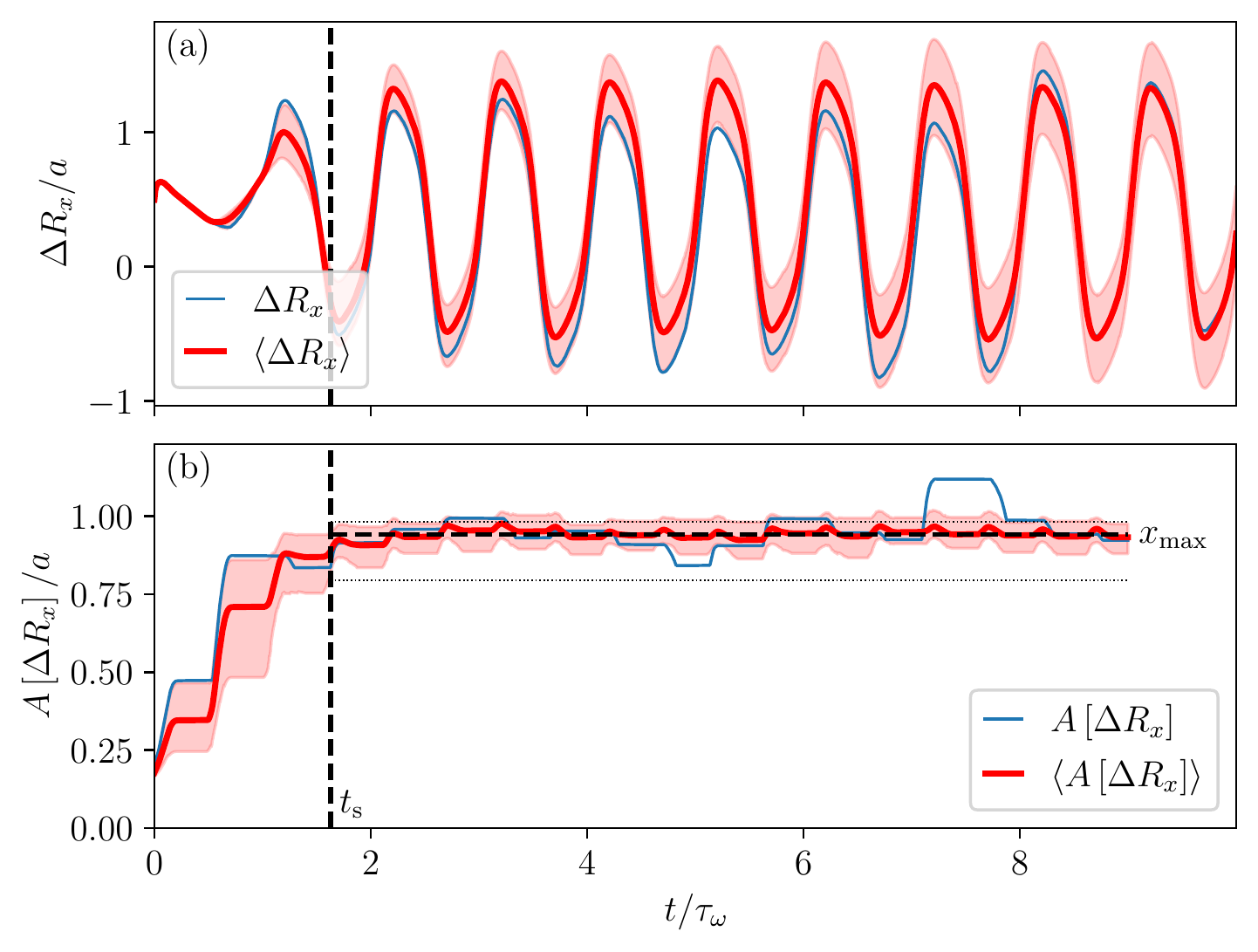}
	\caption{
		(a) Example trajectory of the relative layer motion $\Delta R_x(t)$ in shear-direction and (b) its amplitude $\amp{\Delta R_x(t)}$ at $\dot{\gamma}_0 \tau_\B = 314, \omega \tau_\B = 91.6$. 
		The bold, red line in each graph denotes the ensemble average and the surrounding shaded area the fluctuation range. 
		The thin, blue line represents an example trajectory for a single realization. 
		The average oscillation amplitude increases over the first couple of oscillation periods and saturates at a constant value $x_{\max}$ within a confidence interval (dotted horizontal lines) after the settling time $t_\s$ (dashed vertical line). 
		The saturated amplitude is then taken as the time-average between $[t_\s, T]$ (dashed horizontal line), $T=10 \tau_\omega$ being the total simulation time.
	}
	\label{fig:sample_oscillation}
\end{figure}

We now proceed towards the case of finite driving frequencies ($\omega > 0$), focusing first on the layer motion.
Due to the time-dependence of the applied shear rate, the system never reaches a non-equilibrium \textit{steady} state. 
Instead, we observe an oscillatory motion of the center-of-mass of each layer that becomes (in the ensemble average) periodic in the long-time limit.
In this situation, one has $\av{\Delta R_x(t)} = \av{\Delta R_x(t + \tau_\omega)}$, where $\Delta R_x$ is the $x$-component of the relative layer vector defined in \eq\eqref{eq:dR} and $\tau_\omega$ is the oscillation period of the shear protocol.
Exemplary results for $\Delta R_x(t)$ are shown in \fig\ref{fig:sample_oscillation}a.

For a given time interval (here usually $\tau_\omega$), we can determine whether the particle layers are pinned to each other or not.
To this end we monitor the oscillation amplitude of the relative center-of-mass position $\amp{\Delta R_x(t)}$ (see \fig\ref{fig:sample_oscillation}b), whose ensemble average saturates at a constant value $x_{\max}= \lim_{t \rightarrow \infty} \av{\amp{\Delta R_x(t)}}$ at long times (for details of the numerical evaluation, see \app\ref{app:amplitude}).
The characterization of depinning via $x_{\max}$ is in analogy to our investigation of the effective single-particle problem, see \se\ref{sec:depinning_single_particle}.
We consider the system as (constantly) pinned if $x_{\max} < a/2$, $a \approx 1.034d$ being the lattice constant of the equilibrium square lattice, because then particles are (on average) not moving past neighbouring particles of the other layer. 
In contrast, if particles do move past them, the system is temporarily depinned and $x_{\max} \geq a/2$.

\begin{figure}
	\includegraphics[width=\linewidth]
	{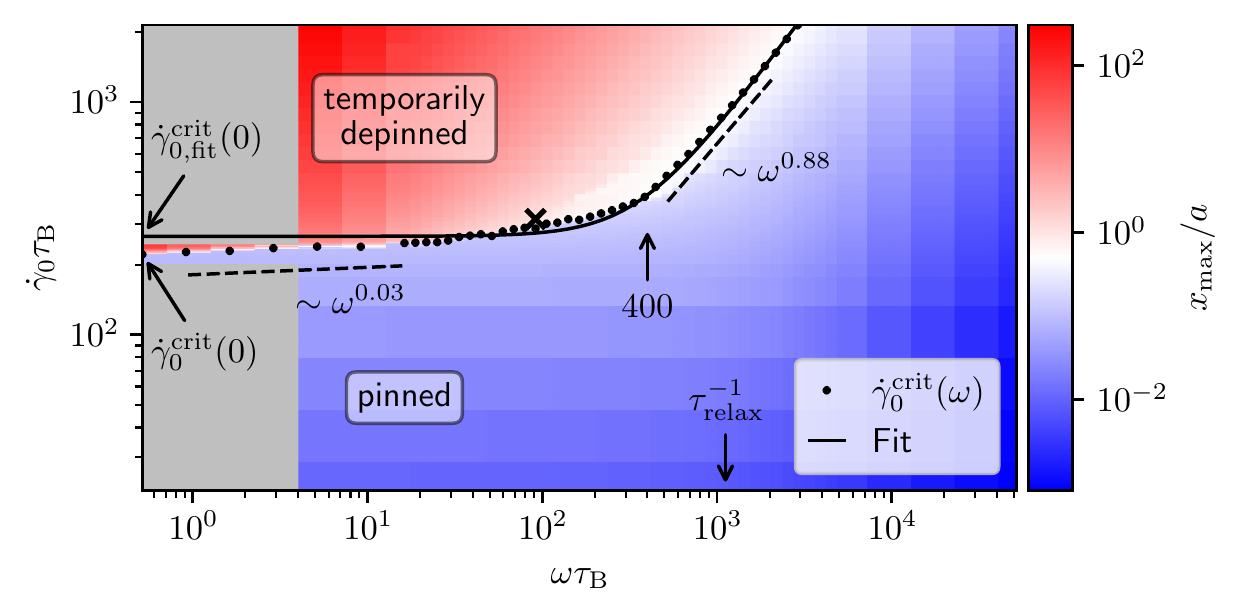}
	\caption{
		Depinning state diagram of the (stochastic) many-particle system in the $(\dot{\gamma}_0, \omega)$ parameter space. 
		Analogously to the single-particle case (compare \fig\ref{fig:sd_single_particle}), blue (red) regions represent constantly pinned (temporarily depinned) states.
		Grey regions have not been computed.
		The critical shear rate amplitude $\dot{\gamma}_0^\crit(\omega)$ (dots) follows the contour line $x_{\max} = a/2$ and is fitted by \eq\eqref{eq:y0crit} (solid line).
		The dashed lines are guides to the eye, indicating the scaling with $\omega$ at very small and very large frequencies.
		The cross within the diagram refers to the parameter combination shown in \fig\ref{fig:sample_oscillation}.
	}
	\label{fig:sd_depinning_many}
\end{figure}

By computing the value of $x_{\max}$ for a mesh of $(\dot{\gamma}_0, \omega)$-parameter combinations, we are able to construct a depinning state diagram, which is shown in \fig\ref{fig:sd_depinning_many} 
in the form of a heatmap. 
Similar to (temporary) depinning in the single-particle problem (see \se\ref{sec:depinning_single_particle}), pinning is generally observed at small $\dot{\gamma}_0$ (blue color) and depinning at large $\dot{\gamma}_0$ (red color).
The transition from pinning to depinning (where $x_{\max}(\dot{\gamma}_0, \omega) \approx a/2$, white color) can be characterized, similar to the case of constant shearing \cite{gerloff2016depinning}, by a critical shear rate amplitude $\dot{\gamma}_0^\crit(\omega)$, which now exhibits a frequency dependence.

In the static limit, the critical amplitude required to induce depinning is $\dot{\gamma}_0^\crit(\omega = 0) \approx \ycrit \tau_\B^{-1}$ \cite{gerloff2016depinning}.
From \fig\ref{fig:sd_depinning_many} we see that this value is approached from above when we decrease $\omega$ from finite values to zero.
Note that $\omega \tau_\B = 0.52$ is the smallest frequency at which we were able to compute $\dot{\gamma}_0^\crit$.
There, it assumes the value $\ycritZero \tau_\B^{-1}$ (labeled as $\dot{\gamma}_0^\crit(0)$ in \fig\ref{fig:sd_depinning_many}).

Upon increasing $\omega$, $\dot{\gamma}_0^\crit(\omega)$ first increases only slightly (with exponent 0.03) and later, for $\omega \tau_\B \gtrsim 400$, almost linearly (with exponent 0.88).
Below this crossover frequency, the transition from pinned to depinned states happens suddenly, indicated by a very slim (white) transition area.
At larger frequencies, in contrast, this transition happens smoothly, indicated by a broader transition area.

Overall, one recognizes a strong similarity between the state diagrams of the shear-driven colloidal bilayer and the driven single particle (\fig\ref{fig:sd_single_particle}) discussed in \se\ref{sec:depinning_single_particle}, with the shear rate amplitude essentially replacing the amplitude of the oscillatory drive.
This already shows that the single-particle model captures important dynamical features of our many-particle system, as long as we focus on the layer motion alone.

Motivated by these analogies, we find that the functional form of $\dot{\gamma}_0^\crit(\omega)$ in the bilayer can be fitted by the function
\begin{equation} \label{eq:y0crit}
\dot{\gamma}_0^\crit(\omega) = \dot{\gamma}_{0, \fit}^\crit(0) \sqrt{1 + \lr{\pi \omega \tau_\rlx}^2} ,
\end{equation}
which provides us with two fit parameters, the critical amplitude in the zero-frequency limit $\dot{\gamma}_{0, \fit}^\crit(0) \approx \ycritZeroFit \tau_\B^{-1}$ and the relaxation time $\tau_\rlx \approx (9.0 \pm 0.2) \cdot 10^{-4} \tau_\B$.
Please note that we deliberately didn't fix $\dot{\gamma}_{0, \fit}^\crit(0)$ to the already known value $\ycrit \tau_\B^{-1}$ \cite{gerloff2016depinning}.
As a consequence, the fitted value overestimates the true zero-frequency limit by about 20\%.
Interestingly however, it approximately matches the transition shear rate from disordered to hexagonal steady states $\dot{\gamma}_0^{\crit,2} \tau_\B = \ycritB$ in the case of constant shearing.
Further, $\tau_\rlx$ can be interpreted as the many-particle equivalent of the substrate relaxation time in the effective single-particle model [see \eq\eqref{eq:tau_sub}]. 
This is also approximately the time that a shear-distorted square lattice ($\dot{\gamma}_0 \tau_\B < \ycrit$) would need to return to its equilibrium state if the external shearing was suddenly stopped.
\subsection{Settling time at the depinning transition} \label{sec:settling_time}
So far we have discussed the existence of temporary depinning based on oscillation amplitudes $x_{\max}$ at long times.
Another interesting quantity is the settling time $t_\s$, which is the time span between the onset of oscillatory shear and the stabilization of $x_{\max}$ (for a visualization, see \fig\ref{fig:sample_oscillation}b).
A detailed definition of $t_\s$ can be found in \app\ref{app:settling_time}.
Note that we start shearing from an equilibrium square lattice.

\begin{figure}
	\includegraphics[width=\linewidth]
	{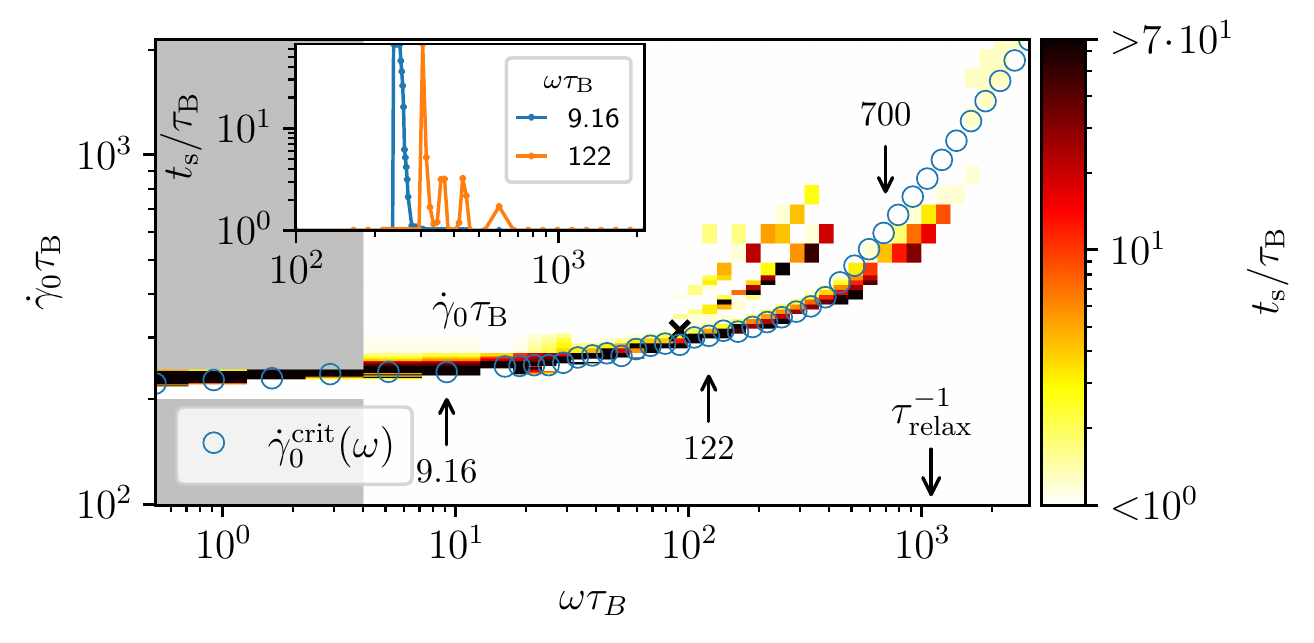}
	\caption{
		Settling time heatmap $t_\s(\dot{\gamma}_0, \omega)$. 
		For the majority of parameter combinations $(\dot{\gamma}_0, \omega)$, the layer's oscillatory movement has settled within one oscillation period $\tau_\omega$ (white color). 
		For shear rate amplitudes $\dot{\gamma}_0$ closely above the critical amplitude $\dot{\gamma}_0^\crit(\omega)$ (blue circles), the settling time increases drastically (yellow to black color).
		The inset shows two slices $t_\s(\dot{\gamma}_0)$ for constant $\omega$.
		Some frequencies are highlighted by arrows.
		The cross within the diagram refers to the parameter combination shown in \fig\ref{fig:sample_oscillation}.
	}
	\label{fig:sd_settling_time}
\end{figure}

From a physical point of view, the settling time can be considered as another relevant time scale that interferes with the relaxation time $\tau_\rlx$, introduced in \eq\eqref{eq:y0crit}.
Results for $t_\s(\dot{\gamma}_0, \omega)$ are shown in \fig\ref{fig:sd_settling_time}.
For the majority of $(\dot{\gamma}_0, \omega)$ combinations, the settling time adopts values below one oscillation period (i.e. $t_\s < \tau_\omega$), which cannot be resolved. Hence, the long-time limit is reached almost instantaneously. 
In contrast, for $(\dot{\gamma}_0, \omega)$ combinations closely above $\dot{\gamma}_0^\crit(\omega)$, $t_\s$ increases drastically, oftentimes even beyond $67 \tau_\omega$, which is the upper measurable limit we set for this work.
The steep left flank of the function $t_\s(\dot{\gamma}_0)$ (compare the inset in \fig\ref{fig:sd_settling_time}) indicates that the settling time might even diverge upon approaching $\dot{\gamma}_0^\crit(\omega)$.
This divergent behaviour disappears, however, at frequencies beyond $\omega \tau_\B > 700$.
Above this frequency the area of increased $t_\s$ starts to deviate from $\dot{\gamma}_0^\crit(\omega)$ (towards smaller $\dot{\gamma}_0$).
For even larger frequencies, the settling time vanishes completely beyond $\omega \tau_\B > 1400$.

Note that the deviation from $\dot{\gamma}_0^\crit(\omega)$ at large $\omega$ is connected to a shift of the particles' oscillation center, as well as a change of the lateral structure (see \se\ref{sec:structure}).
Hereby, the particles tend to oscillate around the lattice sites of their adjacent layer particles instead of their own, analogously to the behaviour seen in the effective single-particle problem (for details, see \app\ref{app:xmean}).
Another striking similarity to the single-particle case (compare \fig\ref{fig:sd_xmean_single_particle}) is the fact that there are two additional small bands with increased settling time above $\dot{\gamma}_0^\crit(\omega)$ in the range $\omega \tau_\B \in [100, 400]$.
\subsection{Structural transitions \label{sec:structure}}
Until now we have explored the center-of-mass motion of the oscillating layers and, based on that, the depinning transition.
We now turn to the analysis of structural phenomena within the layers that accompany the depinning.
We note that our focus here lies on the \textit{average} type of structure, an investigation of heterogeneities is beyond the scope of this study.

\begin{figure}
	\centering
	\subfloat
		[
			square $\square$ \hspace{\textwidth}
			$\psi_4 = 0.99$ \hspace{\textwidth}
			$\psi_6 = 0.08$
		]
		{
			\label{fig:snapshot_square}%
			\includegraphics[width=0.32\linewidth]
			{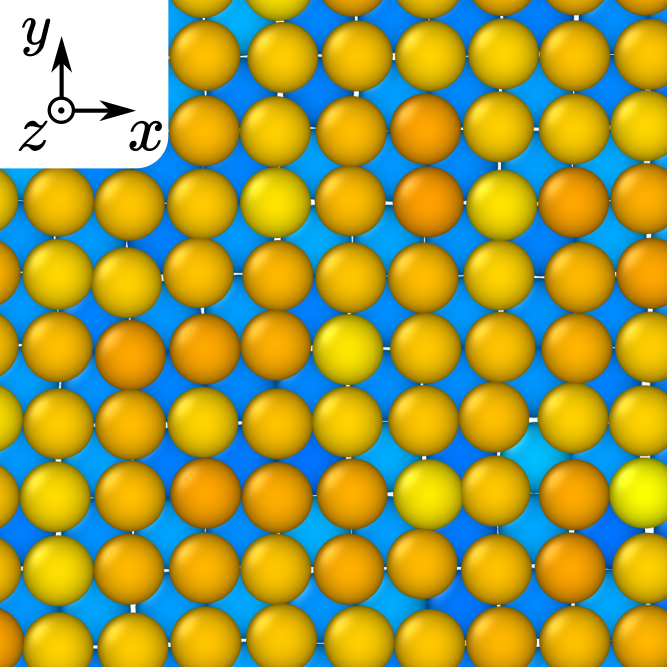}%
		}
	\hfill
	\subfloat
		[
			disordered $\times$ \hspace{\textwidth}
			$\psi_4 = 0.13$ \hspace{\textwidth}
			$\psi_6 = 0.58$
		]
		{
			\label{fig:snapshot_disordered}%
			\includegraphics[width=0.32\linewidth]
			{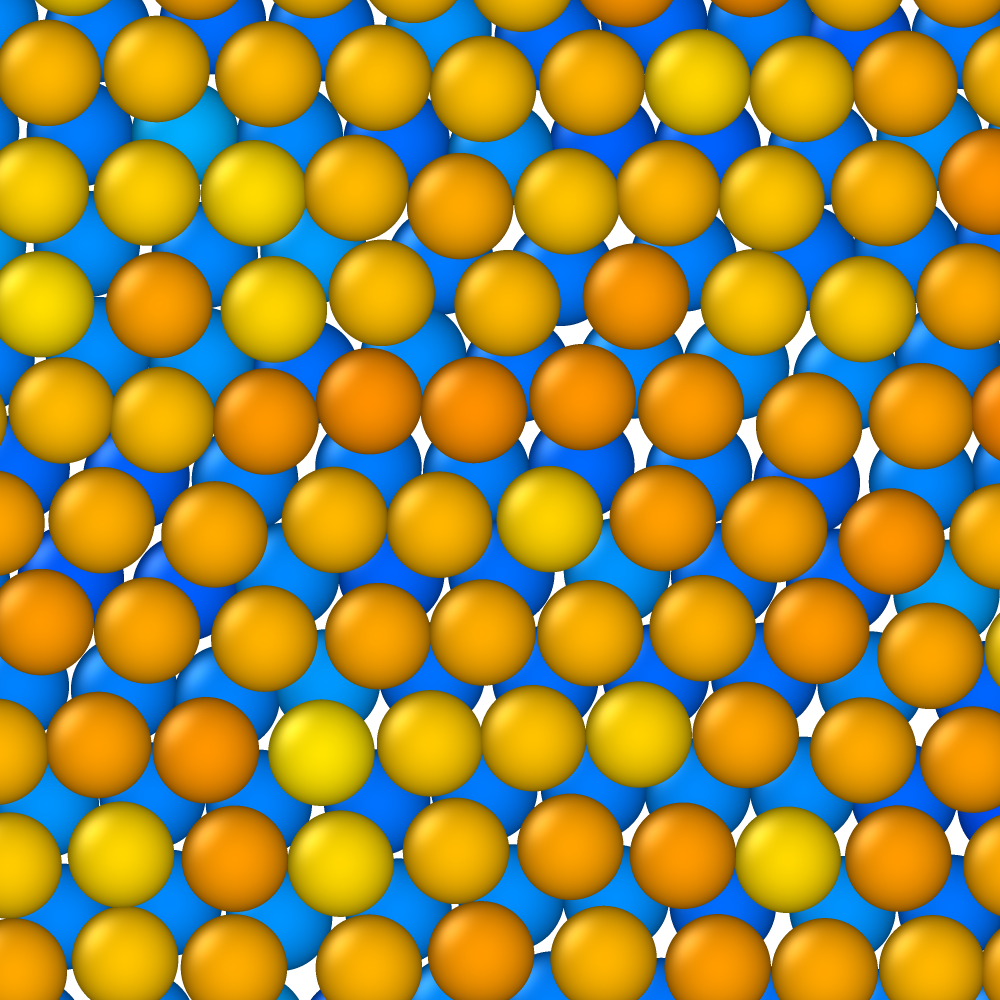}%
		}
	\hfill
	\subfloat
		[
			hexagonal $\hexagon$ \hspace{\textwidth}
			$\psi_4 = 0.16$ \hspace{\textwidth}
			$\psi_6 = 0.87$
		]
		{
			\label{fig:snapshot_hexagonal}%
			\includegraphics[width=0.32\linewidth]
			{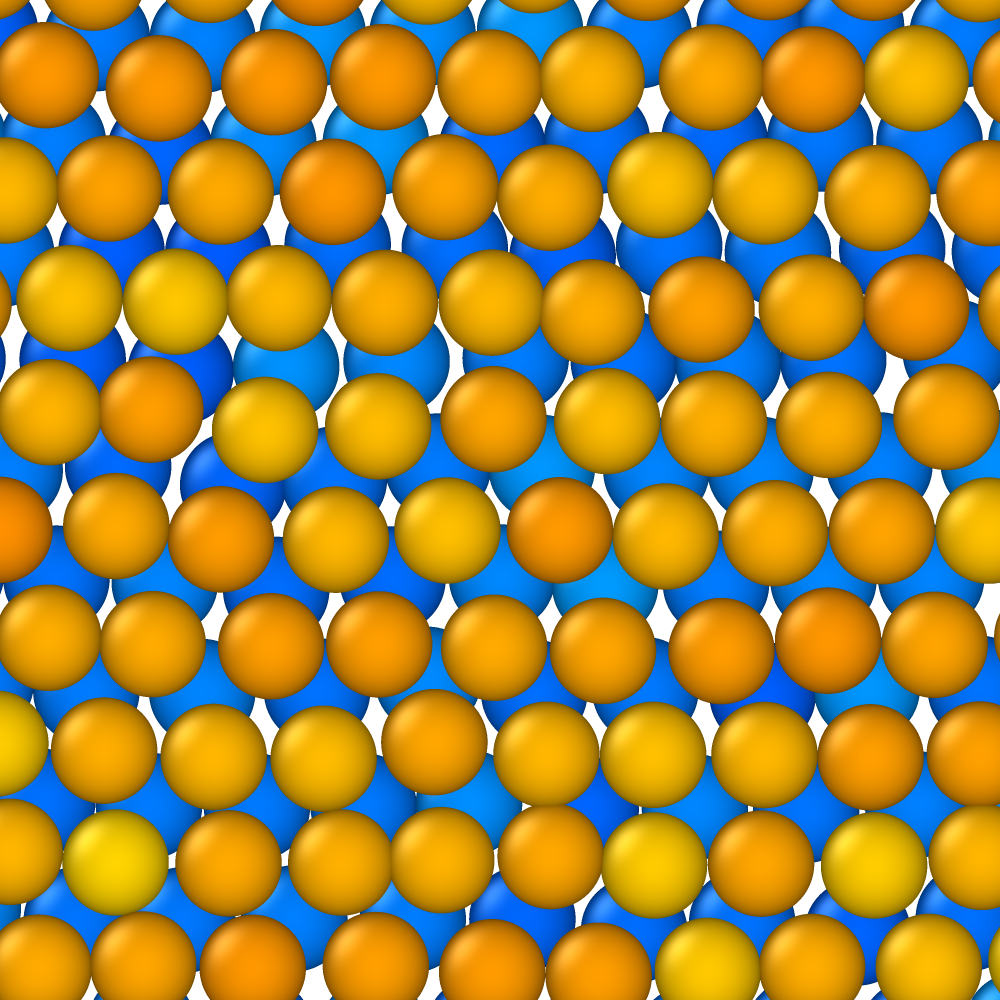}%
		}
	\caption{
		Top-view structure snapshots of a system with $\dot{\gamma}_0 \tau_\B = 769, \omega \tau_\B = 51.5$ (temporary depinning) at  (a) $t=0$, (b) $t=2.8 \tau_\omega$ and (c) $t=3.1 \tau_\omega$. 
		Particles in the upper and lower layer are colored yellow and blue, respectively. 
		The corresponding time dependence of order parameters is given in \fig\ref{fig:angBond_example}. 
		Rendered with OVITO \cite{stukowski2009visualization}.
	}
	\label{fig:snapshots}
\end{figure}

\begin{figure}
	\includegraphics[width=\linewidth]
	{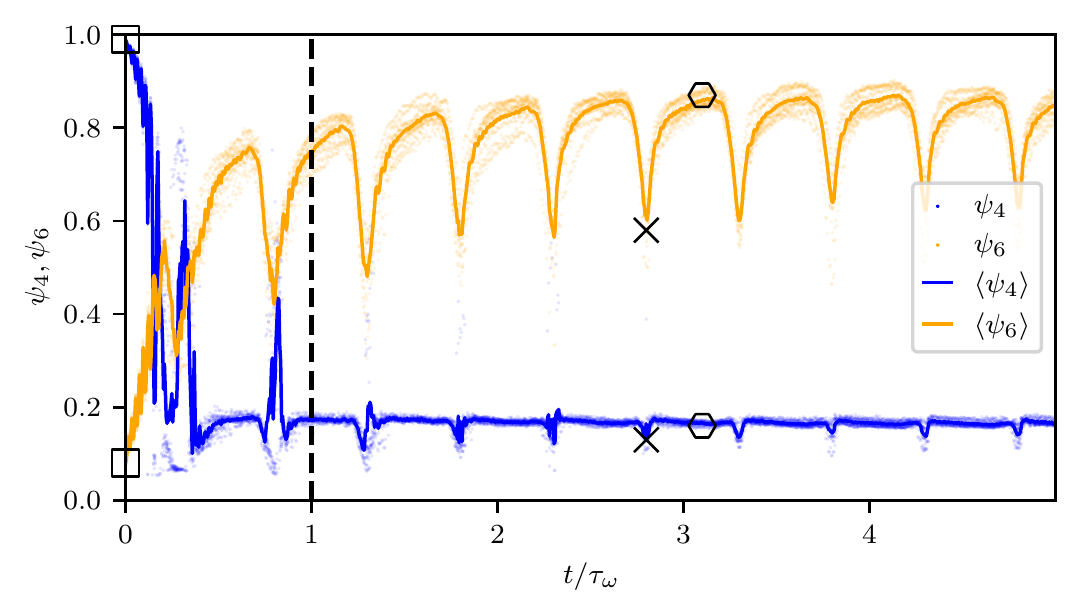}
	\caption{
		Example trajectories of the angular bond order parameters $\psi_4(t)$ (blue) and $\psi_6(t)$ (orange) in a (long-time) disordered-hexagonal state at $\dot{\gamma}_0 \tau_\B = 769, \omega \tau_\B = 51.5$. Dots represent data from 10 different ensembles and bold lines their ensemble average. The dashed vertical line at $t = 1 \tau_\omega$ indicates the settling time referring to depinning. 
		The oscillation of the angular bond curves does not settle before $t \approx 3 \tau_\omega$.
		The markers (squares, crosses, hexagons) refer to the snapshots shown in \fig\ref{fig:snapshots}.
	}
	\label{fig:angBond_example}
\end{figure}

For static shear \cite{vezirov2013nonequilibrium, gerloff2016depinning}, the layers of the present bilayer system can adopt three kinds of structures: square (s), disordered (d) and hexagonal (h). 
All of these structures are also encountered when applying oscillatory shear.
For an illustration in presence of depinning, see snapshots in \fig\ref{fig:snapshots}.
We characterize these structures qualitatively by computing angular bond order parameters $\psi_n(t)$ (details in \se\ref{sec:quantities_of_interest}) for the symmetries $n=4$ (square) and $n=6$ (hexagonal).
An example of the time dependence of the $\psi_n(t)$ is shown in \fig\ref{fig:angBond_example} for a (long-time) disordered-hexagonal state.
For details of the classification of ``states'', see \app\ref{app:classification}.

Similar to the layer motion (see \fig\ref{fig:sample_oscillation}), the angular bond order parameters develop an (on average) periodic orbit at long times after an initial settling time (see \fig\ref{fig:angBond_example}).
Note that the corresponding ``structural'' settling time is usually larger than the depinning settling time.
Further note that the period of the settled angular bond orbits is half of that of the applied shear rate ($\av{\psi_n(t)} = \av{\psi_n(t+\frac{\tau_\omega}{2})}$), because the formation of a structure only depends on the modulus of the shear rate ($\propto |\cos(\omega t)|$).

\begin{figure}
	\includegraphics
	[width=\linewidth]		
	{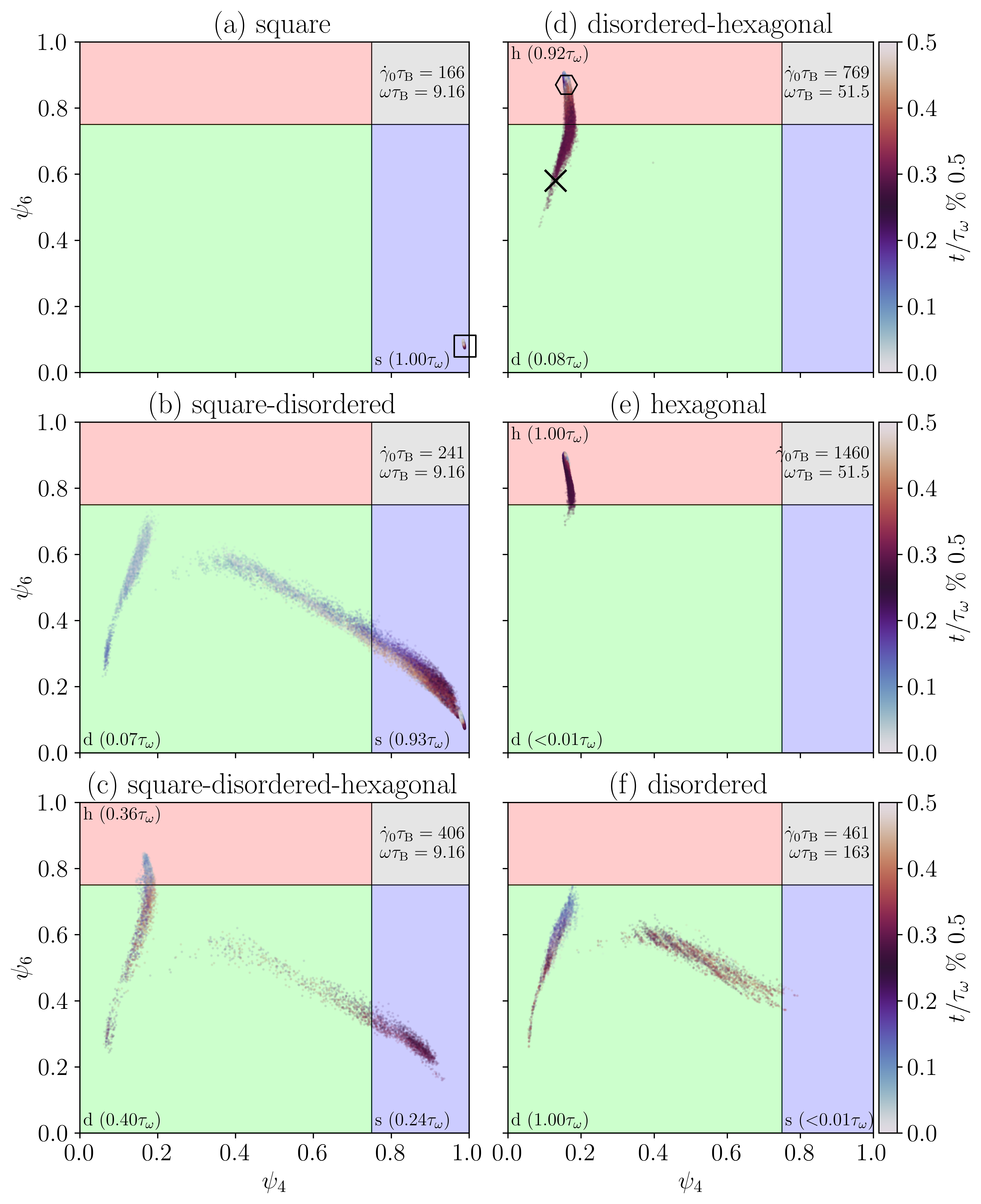}
	\caption{
		Exemplary long-time parametric $(\psi_4, \psi_6)$ curves of the (a) square, (b) square-disordered, (c) square-disordered-hexagonal, (d) disordered-hexagonal, (e) hexagonal and (f) disordered dynamical states.
		The four colored sectors represent the structures: square (blue), disordered (green), hexagonal (red) and forbidden (gray).
		The numbers in parenthesis indicate the amount of time, the particle layers spend in the respective structure during a full oscillation period $\tau_\omega$. 
		The coloring within the curves is mapped onto the current time point within one half-period $\tau_\omega/2$, indicating the orbit orientation.
		Per diagram, every curve consists of 10 ensembles plotted on top of each other, each for at least 3 oscillation periods (6 half-periods).
		The markers (square, cross, hexagon) refer to the snapshots shown in \fig\ref{fig:snapshots}.
	}
	\label{fig:traj_transition_states}
\end{figure}

For the remainder of this section, we focus on the long-time angular orbits.
For this situation we can construct a classification scheme (see \app\ref{app:classification}), which allows us to relate the layer structure at each time to one of the three structural classes (s/d/h).
For an illustration of the different behaviours, it is helpful to look at parametric $(\psi_4, \psi_6)$ curves, for examples see \fig\ref{fig:traj_transition_states}. 
At each timestep, a point is drawn in the $(\psi_4, \psi_6)$ phase space, colored based on the current time within each half-period $\tau_\omega/2$.
During one such half-period, a parametric curve may be located in only one sector of the $(\psi_4, \psi_6)$ plane or in multiple sectors.
We refer to these situations as ``pure'' or ``mixed'' (dynamical) states, respectively (details in \app\ref{app:transition_classification}).
For example, \fig\ref{fig:angBond_example} refers to a mixed disordered-hexagonal (dh) state.

To proceed, we recapitulate the constant shear case, which is recovered formally at $\omega=0$.
Here, we observe pure square states for $\dot{\gamma}_0 \tau_\B < \ycrit$, pure disordered states between $\ycrit \leq \dot{\gamma}_0 \tau_\B < \ycritB$ and pure hexagonal states for $\dot{\gamma}_0 \tau_\B \geq \ycritB$ \cite{vezirov2013nonequilibrium}.
Note that in this case ($\omega = 0$) one may speak of non-equilibrium ``steady'' states because the order parameters indeed become constant (at $t \rightarrow \infty$).

For small but nonzero frequencies $\omega \rightarrow 0$ (for instance $\omega \tau_\B = 9.16$), we instead observe mixed states because the absolute value of the shear rate $\dot{\gamma}(t) = \dot{\gamma}_0 \cos(\omega t)$ crosses the critical thresholds $\dot{\gamma}_0^\crit \tau_\B = \ycrit$ and $\dot{\gamma}_0^{\crit,2} \tau_\B = \ycritB$ during each period, if $\dot{\gamma}_0$ is large enough.
Specifically, for $\dot{\gamma}_0 \tau_\B < \ycrit$ we still observe pure square states (\fig\ref{fig:traj_transition_states}a), for $\ycrit \leq \dot{\gamma}_0 \tau_\B < \ycritB$ mixed square-disordered (sd) states (\fig\ref{fig:traj_transition_states}b), and for $\dot{\gamma}_0 \tau_\B \geq \ycritB$ mixed square-disordered-hexagonal (sdh) states (\fig\ref{fig:traj_transition_states}c).

Upon increase of $\omega$, one would expect to observe the same three dynamical states, but at higher critical thresholds, similar to the increase of the critical depinning shear rate with $\omega$ [compare \eq\eqref{eq:y0crit}].
Although we do indeed observe an increase of these thresholds with $\omega$, we also observe that, at sufficiently large $\dot{\gamma}_0$ (and $\omega$), the system stops returning to the square state.
Instead it forms a disordered-hexagonal (dh) state (\fig\ref{fig:traj_transition_states}d). 
Eventually, it even stops assuming the disordered structure, forming a pure hexagonal state (\fig\ref{fig:traj_transition_states}e).
The latter scenario can be seen, for instance, at $\omega \tau_\B \approx 51.5$, where the particle layers assume the following dynamical states (with increasing $\dot{\gamma}_0$): s ($\dot{\gamma}_0 \tau_\B < 265$), sd ($\dot{\gamma}_0 \tau_\B < 380$), sdh ($\dot{\gamma}_0 \tau_\B < 600$), dh ($\dot{\gamma}_0 \tau_\B < 1000$), h ($\dot{\gamma}_0 \tau_\B \geq 1000$).

\begin{figure}
	\includegraphics[width=\linewidth]
	{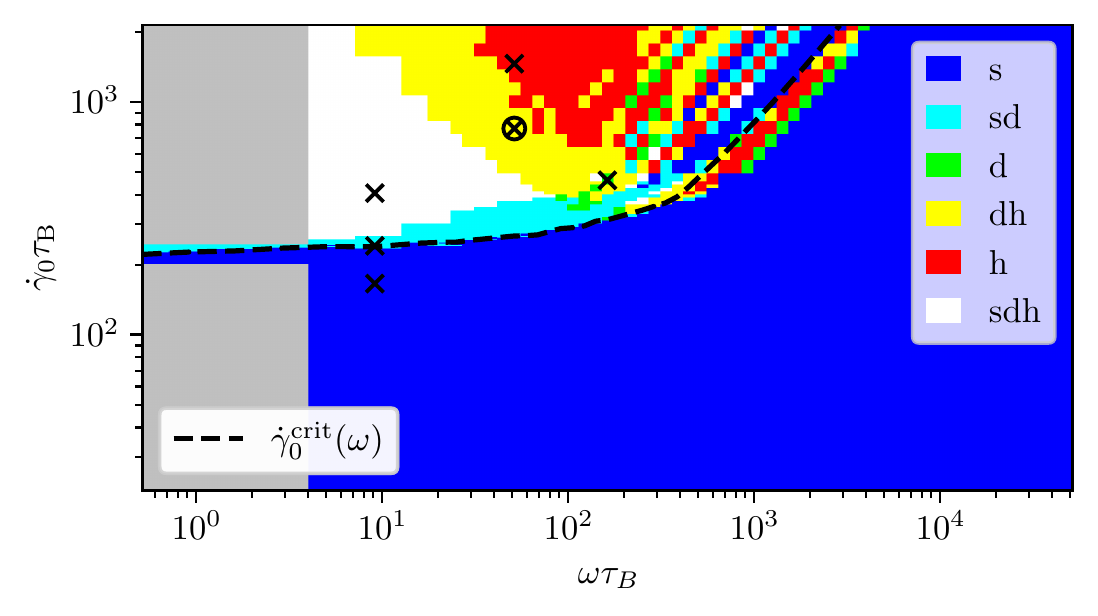}
	\caption{
		Dynamical state diagram of the bilayer in the $(\dot{\gamma}_0, \omega)$ parameter space. The observed dynamical states are square (s), square-disordered (sd), disordered (d), disordered-hexagonal (dh), hexagonal (h) and square-disordered-hexagonal (sdh).
		Markers within the diagram refer to the parameter combinations shown in \fig\ref{fig:angBond_example} (circle) and \fig\ref{fig:traj_transition_states} (crosses).
		The critical (depinning) shear rate amplitude (dots in \fig\ref{fig:sd_depinning_many}) is drawn for reference as dashed, black line.
	}
	\label{fig:sd_dynamical_states}
\end{figure}

Performing simulations for a broad range of combinations $(\dot{\gamma}_0, \omega)$, we can construct a dynamical state diagram, which is shown in \fig\ref{fig:sd_dynamical_states}.
From this diagram, we learn that the aforementioned s-sd-sdh-dh-h sequence (encountered upon increase of $\dot{\gamma}_0$), disappears above $\omega \tau_\B \sim 70$ .
First, the sdh state is dropped, then from $\omega \tau_\B > 120$ onwards, the sequence completely vanishes and we observe all kinds of dynamical states, even purely disordered ones (\fig\ref{fig:traj_transition_states}f) and purely square states at $\dot{\gamma}_0$ well above $\dot{\gamma}_0^\crit(\omega)$.
Interestingly, the alternation between these dynamical states resembles the stability changes of the oscillation centers in the effective single-particle model, discussed briefly in \se\ref{sec:single_particle} and more detailed in \app\ref{app:xmean}.

Finally, we would like to note that the structural details of some dynamical states depends not only on $\dot{\gamma}_0$ and $\omega$, but also on the simulation's system size due to the appearance of heterogeneities in the (microscopic) structure.
Test simulations (see \app\ref{app:system_size}) reveal that the amount of disorder tends to increase for larger systems.
We plan to investigate these dependencies further in a future study.
\section{Conclusions} \label{sec:conclusion}
Based on overdamped Brownian Dynamics simulations we have investigated a dense, colloidal bilayer system under the influence of oscillatory shearing lateral to the confining walls.
Starting from a confinement-induced equilibrium structure characterized by square-like order and investigating a broad range of shear rate amplitudes and frequencies, we have observed and analyzed a variety of dynamical behaviours.

First, we find that the two layers temporarily depin from each other above a frequency-dependent critical amplitude of the applied shear rate, thereby generalizing the behaviour seen in the case of static shearing \cite{vezirov2013nonequilibrium, gerloff2016depinning}.
By comparison with an effective single-particle model for the center-of-mass dynamics of a layer, we have obtained an explicit expression to describe this frequency dependence.
This expression involves a relaxation time, which gives an estimate for a shear relaxation time.

Second, full depinning is reached only after a settling time, which drastically increases near the transition between pinning and depinning (and is thus also frequency-dependent).
This shows that system-intrinsic time scales indeed interact with the externally applied shear frequency, as one would expect.

Third, the (temporary) depinning is accompanied by complex dynamical changes between different in-plane structures, which we refer to as dynamical structures (or states).
Which kinds of dynamical structures form, again depends strongly on both shear parameters, amplitude and frequency.
At slow driving (i.e., small frequencies), these structures are stable against shear parameter changes across multiple orders of magnitude.
At fast driving, on the other hand, small changes of either amplitude or frequency have strong impact on the resulting dynamical state.
Interestingly, in the latter case it is possible to induce non-equilibrium ordered and disordered structures in the layers, while the amplitude of the center-of-mass oscillation remains very small.

Our results can, in principle, be tested for real colloidal systems.
In fact, experiments on confined colloidal layers under shear have been performed, e.g. in \reference\cite{reinmuller2013confined} for film-like systems and Refs.~\cite{ortiz2018laning, gerloff2020dynamical, williams2022rheology} for systems in a circular geometry.
So far, these experiments have been conducted under static shear.
However, particularly the circular geometry may also be suitable to investigate oscillatory shear.%

We note, however, that the predictive power of our simulations might be somewhat limited, since we have neglected the impact of hydrodynamic interactions.
A similar observation was made in the context of a circular sheared system \cite{ortiz2018laning, gerloff2020dynamical}.
As we have shown in the present study that (relaxation) time scales do play an important role under oscillatory shearing, we expect that hydrodynamic interactions might have a stronger impact than under static shear. This becomes especially important at larger shearing frequencies, where the oscillation period starts interacting with the relaxation time scales.
However, including hydrodynamic interactions in the simulations would require an explicit treatment of the flow field of the surrounding solvent, which strongly increases the computational cost.
One major problem in this direction is the treatment of the solid-fluid interface.
Recently, promising advances in the treatment of such have been made possible, e.g. in the framework of the smoothed profile (SP) method \cite{yamamoto2021smoothed}.
Other common approaches to include hydrodynamic interactions in particle-based simulations are, for example, the frameworks of dissipative particle dynamics (DPD), multi-particle collision dynamics (MPCD) or Stokesian dynamics (SD) \cite{park2016review, schiller2018mesoscopic}.

Despite the above mentioned restriction to bilayers, this study gives a flavor of what kind of intriguing behaviour could arise in strongly confined colloidal dispersions under oscillatory shear.
Indeed, we would expect that our findings also apply (at least, qualitatively) to multilayered systems with, e.g., three or four layers.
This expectation foots on an earlier investigation by one of us, where we studied the depinning behavior of such multilayered systems under static shear \cite{gerloff2017shear}.
Specifically, the concept of a frequency-dependent critical shear rate should apply analogously, presumably with a similar course at small and large frequencies.
Larger differences are to be expected regarding the crossing between small and large frequency regime.
Likewise, the concept of dynamical states should also apply to multilayered systems, probably with more complex mixed states at small frequencies and stabilization of pure states at large frequencies.
Indeed, it would be an interesting question, whether all pure states, for example a laned state in a related 3-layer system \cite{gerloff2017shear}, can be stabilized (and pinned) with oscillatory shearing.
 
Based on the variety of newfound structural dynamics under oscillatory shear, we plan to take a closer look at the rheology of these strongly confined, layered systems next, including the analysis of stress-strain relations and the calculation of shear moduli (see \reference\cite{brader2010nonlinearresponse} for a related study in three dimensions).
It has already been shown that these confined systems possess a non-linear constitutive flow curve for static shearing \cite{gerloff2016depinning}.
We thus expect complex stress-strain relations also under oscillatory driving.

In parallel to the (macroscopic) rheological properties, it would also be very interesting to analyze the occurrence and dynamics of spatial heterogeneities (i.e., defects) under oscillatory shear.
In fact, preliminary investigations revealed that these occur quite frequently in the vicinity of depinning, especially when considering larger system sizes.
Related to this, it seems very promising to look at local (non-affine) deformations, which have been intensely investigated in amorphous solids under shear \cite{yeh2020glass, teich2021crystalline, lamp2022brittle}.
Work in these directions is under way.
\section*{Acknowledgements}
We gratefully acknowledge the support of the Deutsche Forschungsgemeinschaft (DFG, German Research Foundation), project number 163436311 - SFB 910 and the HPC cluster of the Institute of Mathematics for computational resources.
We would also like to thank Matthias Wolfrum for useful discussions.
\section*{Source code}
The codes for the many- and single-particle BD simulations are written in C++.
Analysis and visualization tools are implemented in a Python library.
The source code can be provided by the authors upon request.
\appendix
\section{Technical details} \label{app:numerical_details}
\subsection{Finite time step size} \label{app:dt}
In this work, we solve \eq\eqref{eq:overdamped-sde} numerically for different parameter combinations of $\dot{\gamma}_0$ and $\omega$.
We therefore have to choose the discretized time step size $\Delta t$, which enters via the discretization of $\dot{\vec{r}}_i(t) \approx (\vec{r}_i(t+\Delta t) - \vec{r}_i(t))/\Delta t$, carefully. 
Following \reference\cite{wang2016large}, we first determine the relevant time scales in the system and then choose the time step size as a fraction ($\epsilon = 10^{-3}$) of the smallest of them, $\Delta t = \epsilon \min \lr{\tau_\omega, \tau_\mathrm{d}, \tau_\B, \tau_\mathrm{relax}}$.
Regarding the time scales, we consider the oscillation period $\tau_\omega = 2\pi/\omega$, the deformation time $\tau_\mathrm{d} = 1/\dot{\gamma}_0$, the Brownian time $\tau_\B = d^2/D_0$ and an estimated relaxation time $\tau_\mathrm{relax} \approx 10^{-2} \tau_\B$.
Note that the estimation of $\tau_\mathrm{relax}$ is not trivial because it strongly depends on the stiffness of the particle's surrounding potential landscape and thus on the underlying model parameters and current state of the system.
Indeed, it turns out that $\tau_\mathrm{relax}$ is overestimated by a factor $10$ (compare \se\ref{sec:temporary_depinning}). 
Nonetheless, in conjunction with the multiplication by $\epsilon$ (which compensates this error), we end up with a maximal time step size of $\Delta t \leq 10^{-5} \tau_\B$, which has already been successfully applied in previous studies \cite{vezirov2013nonequilibrium, gerloff2016depinning}.
These time scales refer to the sheared slit-pore many-particle system (\se\ref{sec:model_many_particle}). 

For the effective single-particle model (\se\ref{sec:single_particle}), we use an analogous approach.
Instead of $\dot{\gamma}_0$, we have $F_{\dr,0}$ as driving amplitude, and the equation of motion is \eq\eqref{eq:sde_1d}, which we solve with an Euler method, starting from $x_0=0$, computing a total of 10 oscillation periods.
The above definitions of oscillation period and Brownian time still apply, but the deformation time is replaced by a driving time $\tau_\dr = a/(\mu F_{\dr,0})$ and the (substrate) relaxation time is defined as $\tau_\sub = a/(2\pi \mu F_{\sub,0})$ [see \eq\eqref{eq:tau_sub}].

Note that the time step size $\Delta t$ has to be determined separately for each considered parameter combination of $\dot{\gamma}_0$ (or $F_{\dr,0}$) and $\omega$, since the oscillation period and deformation (or driving) time depend on it.
\subsection{Intra-layer pair correlation function} \label{app:pair_correlation}
In this section, we describe the calculation of the (instantaneous) intra-layer pair correlation function that we use to determine the nearest neighbour distance, which is needed in the definition of angular bond order parameters in \eq\eqref{eq:psi_n}.
For each layer $m=1,2$, the intra-layer pair correlation function at time $t$ and for a specific noise realization is defined by
\begin{equation}
	g^\intra_m(r) = \frac{1}{\rho_m^\tot N_m 2 \pi r} \sum_{i=1}^{N} \sum_{j \neq i}^{N} H_m(z_i) H_m(z_j) \delta(r - r_{ij}) ,
\end{equation}
which contains the area density of particle pairs with distance $r$ [$\sim \delta(r-r_{ij})/(2\pi r)$] compared to the total area density $\rho_m^\tot = N_m/(L_x L_y)$ of all $N_m$ particles within that layer. 
Since the layers are all aligned parallel to the $x$-$y$-plane, only in-plane distances $r_{ij} = \sqrt{x_{ij}^2 + y_{ij}^2}$ are considered (taking $z_{ij}$ into account changes $g^\intra_m(r)$ only marginally). 
Furthermore, the pair of layer identification functions $H_m(z_i) H_m(z_j)$ [see \eq\eqref{eq:layer_id}] ensures that only particle pairs within that layer are counted. 
Numerically, we treat the delta-function $\delta(r)$ as a rectangular pulse of width  $\Delta r = 0.05d$ and height $1/\Delta r$. 
In this representation, the area density can be thought of as the number of particles within an annulus of radius $r$ and width $\Delta r$ ($r_{ij} \in [r - \Delta r/2, r + \Delta r/2]$) divided by its area $2\pi r \Delta r$.

Due to the symmetry of the considered bilayer system with respect to the plane $z=0$, both particle layers exhibit the same structure, besides from temperature fluctuations. 
Thus, it is sufficient to consider only the average across both layers
\begin{equation}
	g^\intra(r) = \frac{1}{N_L} \sum_{m=1}^{N_L} g^\mathrm{intra}_m(r) .
\end{equation}
\subsection{Oscillation amplitude} \label{app:amplitude}
For a function $x(t)$ (here usually a single or averaged particle position), we define the oscillation amplitude $\amp{x(t)}$ at time $t$ as half the distance between the largest and the smallest value in a forward oriented time window of length $\tau_\omega=2\pi/\omega$, 
\begin{equation} \label{eq:windowed_amplitude}
	\amp{x(t)} = \frac{1}{2} \lr{\ceil{x(t')}_{t'=t}^{t'=t+\tau_\omega} - \floor{x(t')}_{t'=t}^{t'=t+\tau_\omega}} .
\end{equation}
Here, we define the maximum and minimum values of $x(t')$ in the interval $t' \in [a, b]$ as
\begin{subequations}
	\begin{align}
		\ceil{x(t')}_{t'=a}^{t'=b} &= \max \llrr{x(t') | a \leq t' < b} , \\
		\floor{x(t')}_{t'=a}^{t'=b} &= \min \llrr{x(t') | a \leq t' < b}.
	\end{align}
\end{subequations}
In our oscillatory driven systems, the oscillation amplitude usually saturates at long times (at least in the ensemble average).
We refer to the constant long-time value as
\begin{subequations} \label{eq:xmax}
\begin{align} 
	x_{\max} 	&= \lim_{t \rightarrow \infty} \av{\amp{x(t)}} \\
				&\approx \av{\amp{x(t_\s)}} \\
				&\approx \frac{1}{T - t_\s} \int\limits_{t_\s}^T \mathrm{d}t \av{\amp{x(t)}} , \label{eq:xmax_time_average}
\end{align}
\end{subequations}
which is approximately already reached at the (finite) settling time $t_\s$ (defined in \app\ref{app:settling_time}).
Note that the amplitudes are calculated before the ensemble averages.
To further reduce remaining noise (despite the ensemble-averaging), we additionally take values for $t < t_\s$ up to the total simulation time $T$ into account by computing the time-average in \eq\eqref{eq:xmax_time_average}.
\subsection{Settling time} \label{app:settling_time}
In \se\ref{sec:settling_time}, we introduced the settling time $t_\s$ of the relative layer motion, which we defined as the time span between the onset of oscillatory shear and the stabilization of the oscillation amplitude $x_{\max}$.
In this section, we explain our definition of $t_\s$, specifically the saturation criterion of $x_{\max}$.

Generally speaking, we consider a time-dependent observable $x(t)$ (e.g. $x=\av{\amp{\Delta R_x}}$), which saturates at $x_\infty = const$ for long times: $x(t \rightarrow \infty) = x_\infty$. 
This provided, we define the settling time $t_\s$ as the earliest point in time, when $x(t) \approx x_\infty$ for the remaining $t \geq t_\s$.
In the following, we specify in more detail, how we implement this saturation criterion for an observable that has been computed for a finite simulation time $t \in [0, T]$. 
First, we define the maximal relative distance between $x(t)$ and any following function value at a later time $t' > t$
\begin{equation} \label{eq:conv_crit}
	\delta(t) = \max \llrr{ 
		\left.\frac{\abs{x(t) - x(t')}}{\abs{x(t)}} \right| t < t' \leq T
	} .
\end{equation}
Now, the saturation condition $x(t) \approx x_\infty, t \geq t_\s$ is mapped onto $\delta(t) \approx 0, t \geq t_\s$.
Note that $\delta(t)$ deliberately contains information from the entire time interval $t' \in (t, T]$, not only from its end ($t' = T$), to capture potential oscillatory behaviour in-between.

Finally, we define the settling time as the smallest time, where $\delta(t)$ becomes smaller than a convergence threshold $\epsilon > 0$
\begin{equation} \label{eq:settling_time}
	t_\s = \min \llrr{t \left| \delta(t) < \epsilon, \tau_\omega \leq t \leq \frac{2}{3} T \right.} .
\end{equation}
Here, we demand that the saturation criterion is fulfilled at least until $t = \frac{3}{2} t_\s$. Thus, only settling times up to an upper limit of $\frac{2}{3}T$ are measurable.
If there is no $t \leq \frac{2}{3}T$, which fulfills the saturation condition $\delta(t)<\epsilon$, we assume that the settling time diverges: $t_\s = \infty$.
Additionally, in our specific case, settling times below one oscillation period cannot be resolved, because the amplitude definition in \eq\eqref{eq:windowed_amplitude} includes information from function values in a time window of length $\tau_\omega$. Hence, we apply the lower limit $t_\s \geq \tau_\omega$.

For our numerical calculations, we choose a convergence threshold of $\epsilon=0.1$. 
However, reasonable results can be obtained in the range $\epsilon \in [0.03, 0.15]$.
Note that the value of $\epsilon$ represents the systematic relative error of the saturated value $x(t_\s)$.
Finally, we compute at least $T=10\tau_\omega$ oscillation periods, increasing the simulation time stepwise (10-30-50-100), if no settling time ($t_\s = \infty$) can be found at first. 
To prevent excessively time- and memory-consuming simulations in the case of diverging settling times, we don't go higher than $T=100\tau_\omega$. 
Similarly, to obtain reasonable ensemble averages, we compute 10-100 realizations.
\section{Oscillations in the strongly driven single-particle model} \label{app:xmean}
In this section, we sketch the derivation of the analytical solution of \eq\eqref{eq:sde_1d} at large driving amplitudes and zero noise, based on Refs.~\cite{chow1985ring} and \cite{cotteverte1994vectorial} on a similar problem in the context of laser physics.
For readability, we state here again the deterministic equation
\begin{equation} \label{eq:ode_1d}
	\dot{x} = - F_{\sub,0} \sin \left( \frac{2\pi}{a} x \right)
	+ F_{\dr,0} \cos(\omega t) .
\end{equation}

For large amplitude motion, the solution to \eq\eqref{eq:ode_1d} will be similar to that of a free particle [compare \eq\eqref{eq:xmaxFree}]. 
Therefore, we make the ansatz
\begin{equation} \label{eq:ansatz_large_w}
	x(t) = \frac{F_{\dr,0}}{\omega} \sin(\omega t) + \theta(t) ,
\end{equation}
where $\theta(t)$ is assumed to be a small correction.
Inserting \eq\eqref{eq:ansatz_large_w} into \eq\eqref{eq:ode_1d} and isolating $\dot{\theta}(t)$ yields
\begin{equation} \label{eq:ode_delta}
	\dot{\theta}(t) = - F_{\sub,0} \sin \lr{ \frac{2\pi}{a} \LR{ \frac{F_{\dr,0}}{\omega} \sin(\omega t) + \theta(t) } } .
\end{equation}
Utilizing a Bessel-Fourier expansion $\sin(x \sin \beta) = \sum_{m=-\infty}^{\infty} J_m(x) \sin(m \beta)$ (following \reference\cite{chow1985ring}) and the trigonometric identity $\sin(\alpha + \beta) = \sin(\alpha) \cos(\beta) + \cos(\alpha) \sin(\alpha)$, we obtain
\begin{equation} \label{eq:ode_delta_bessel}
	\dot{\theta}(t) = - F_{\sub,0} \sum_{m=-\infty}^{\infty} J_m \lr{ \frac{2 \pi F_{\dr,0}}{a \omega} } \sin \lr{ m \omega t + \frac{2 \pi}{a} \theta(t) } ,
\end{equation}
where $J_m(x)$ is the $m$-th Bessel function of the first kind.
So far, \eq\eqref{eq:ode_delta_bessel} is still exact.
Assuming that $\theta(t)$ slowly varies in time compared to $\omega t$, we neglect all summands except the one with $m=0$, yielding
\begin{equation} \label{eq:ode_delta_m0}
	\dot{\theta}(t) \approx - F_{\sub,0} J_0 \lr{ \frac{2 \pi F_{\dr,0}}{a \omega} } \sin \lr{ \frac{2 \pi}{a} \theta(t) } .
\end{equation}
\eq\eqref{eq:ode_delta_m0} represents an Adler-type equation for $\theta(t)$ with the solution \cite{mathematica10}
\begin{equation} \label{eq:theta_solution}
	\theta(t) = \frac{a}{\pi} \arccot \LR{ \cot \lr{ \frac{\pi}{a} x_0 } \exp \lr{ J_0 \lr{ \frac{2 \pi F_{\dr,0}}{a \omega} } \frac{t}{\tau_\sub} } } .
\end{equation}
Here, we inserted the initial condition $x(0)=\theta(0)=x_0$.

\eq\eqref{eq:ode_delta_m0} has two fixed points at $\theta = 0$ and $\theta = a/2$. 
Which one of them is stable depends on the sign of $J_0 \lr{ \frac{2 \pi F_{\dr,0}}{a \omega} }$ and therefore on the combination of $F_{\dr,0}$ and $\omega$.
As a slowly varying correction, $\theta(t)$ describes the time-averaged oscillation center $\pav{x}(t) \approx \theta(t)$ of the (otherwise fast) sinusoidal particle motion, where $\pav{x}(t) = \int_t^{t+\tau_\omega} x(t') \mathrm{d}t'/\tau_\omega$ denotes the average over one period.
The presence of the two fixed points implies that the oscillation center assumes only two possible values, either $\theta = 0$ or $\theta = a/2$ for $t \rightarrow \infty$, corresponding to a stable oscillation around the minimum or maximum of the substrate potential [\eq\eqref{eq:Vsub}], respectively.
Specifically, the particle oscillates around the potential minimum ($\theta = 0$), if $J_0 \lr{ \frac{2 \pi F_{\dr,0}}{a \omega} } > 0$, and around the potential maximum ($\theta = a/2$), if $J_0 \lr{ \frac{2 \pi F_{\dr,0}}{a \omega} } < 0$.

Taking a closer look at the shape of the zeroth order Bessel function $J_0$, we recognize that it exhibits an infinite amount of zero crossings $j_{0,k}$, $k = 1, 2, \dots$, starting from positive values for small arguments [$J_0(0) = 1$].
This means that, at constant $\omega$, the particle's stable oscillation center will periodically swap between potential minimum and maximum, when increasing $F_{\dr,0}$, starting from an oscillation around the potential minimum (small $F_{\dr,0}$).
Moreover, expressed as functions $F_{\dr,0}(\omega)$, the locations of these ``stability swaps'' are given by
\begin{equation} \label{eq:Fswapk}
	F_{\dr,0}^{\swap, k}(\omega) = j_{0,k} F_{\sub,0} \omega \tau_\sub , \quad k = 1, 2, \dots
\end{equation}
The first swap ($k=1$) from $\theta=0$ to $\theta=a/2$ follows from \eq\eqref{eq:Fswapk} with $j_{0,1} \approx 2.4048$.

\begin{figure}
	\includegraphics[width=\linewidth]
	{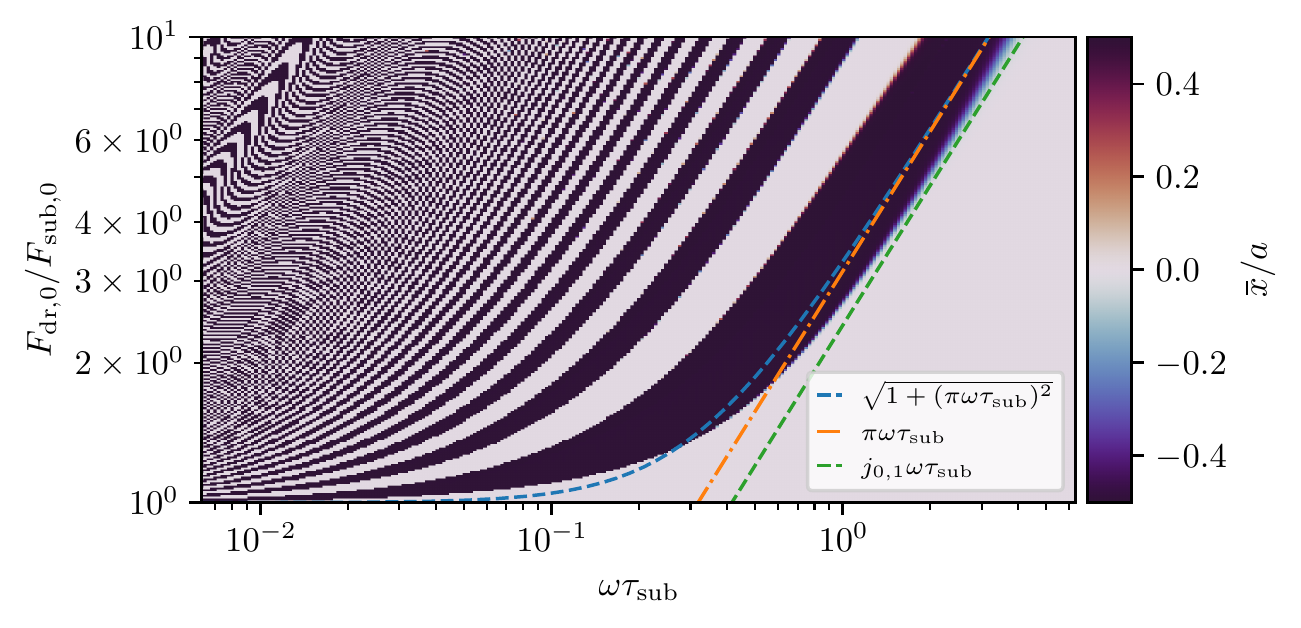}
	\caption{
		Heatmap of the oscillation centers $\pav{x}$ in the deterministic single-particle model.
		Light and dark colored areas indicate solutions, where the particle oscillates around the substrate potential minimum $\pav{x}=0$ and maximum $\pav{x}=a/2$, respectively.
		At large amplitudes (and frequencies), \eq\eqref{eq:Fswapk} describes the position of the first stability swap from $\pav{x}=0$ to $\pav{x}=a/2$ (green dashed line), at slightly smaller driving force amplitudes than $F_{\dr,0}^\crit$ [\eq\eqref{eq:Fdr0_crit}, orange dash-dotted line]. 
		At smaller frequencies (blue dashed line), the transition happens at slightly larger driving force amplitudes instead.
	}
	\label{fig:sd_xmean_single_particle}
\end{figure}

Comparing these predictions with our numerical investigation of the single-particle model (see \fig\ref{fig:sd_xmean_single_particle}), 
we found that the stability of $\theta$ indeed periodically swaps in large portions of the $(F_{\dr,0}, \omega)$ parameter space upon increasing $F_{\dr,0}$ (above $F_{\dr,0} \geq F_{\sub,0}$).
The predicted swapping locations given by \eq\eqref{eq:Fswapk}, however, are only valid at large driving amplitudes ($F_{\dr,0} \gtrsim 3 F_{\sub,0}$) and large frequencies ($\omega \tau_\sub \gtrsim 10^{-1}$).
Note that the first swap from $\theta=0$ to $\theta=a/2$ at large frequencies [\eq\eqref{eq:Fswapk}, $k=1$, dashed green line] already happens below the critical driving amplitude [\eq\eqref{eq:Fdr0_crit}, orange dash-dotted line], at an oscillation amplitude of  $x_{\max} = \frac{j_{0,1} a}{2 \pi} \approx 0.38 a$.
\section{Structural classification details} \label{app:classification}
\subsection{Criteria and threshold values} \label{app:struc_classification}
Our classification is based on the two time-dependent angular bond order parameters $\psi_4$ and $\psi_6$ defined in \eq\eqref{eq:psi_n}, each of them assuming values between zero and one.
Note that the definition of $\psi_4$ and $\psi_6$ [see \eq\eqref{eq:psi_n}] involves an average over all particles, thus this approach ignores any heterogeneities that may occur within or between the particle layers.

Since we are dealing with a noisy system, perfect square or hexagonal structures ($\psi_{4/6} = 1$) are not observed.
Therefore, we choose threshold values $\psi_4^\thr = \psi_6^\thr = 0.75$ to allow for impurities.
Based on these threshold values, we classify a pair of $(\psi_4, \psi_6)$ values into one of four structures: square (s), disordered (d), hexagonal (h) or forbidden (h), see \tab\ref{tab:structure_classification}.
\begin{table}[t]
	\caption{Criteria for the structure classification.}
	\label{tab:structure_classification}
	\begin{tabular}{c|c|c}
		structure		&	abr.	&	criterion	\\
		\hline
		square		&	s		&	$\psi_4 > \psi_4^\thr$, $\psi_6 \leq \psi_6^\thr$	\\
		disordered	&	d		&	$\psi_4 \leq \psi_4^\thr$, $\psi_6 \leq \psi_6^\thr$		\\
		hexagonal	&	h		&	$\psi_4 \leq \psi_4^\thr$, $\psi_6 > \psi_6^\thr$	\\
		forbidden	&	f		&	$\psi_4 > \psi_4^\thr$, $\psi_6 > \psi_6^\thr$
	\end{tabular}
\end{table}
The ``forbidden'' case occurs because the local structure around a particle cannot be, at the \textit{same} time, square \textit{and} hexagonal.
The visual counterpart of this classification is to divide the $(\psi_4, \psi_6)$ plane into four rectangular sectors (see \fig\ref{fig:traj_transition_states}), where each sector is attributed to one structure.

Note that the choice of the threshold values is arbitrary to a certain extent.
We found that threshold values in the range $[0.6, 0.8]$ are appropriate.
Going below $0.6$, one would start to observe ``forbidden'' cases, while going above $0.8$, one would primarily identify disordered structures.
Furthermore, since the angular bond curves feature jumps within the disordered sector (there are some void areas in the phase space, see e.g. \fig\ref{fig:traj_transition_states}f), it is especially important to keep $\psi_6^\thr$ well above these jumping areas.

As a prerequisite for the next section, it is practical to define a structure identifier function $p(\alpha, \psi_4, \psi_6)$ ($\alpha \in \{s,d,h,f\}$), which is one, if the system is in structure $\alpha$, or zero, if it is not, based on the criteria in \tab\ref{tab:structure_classification}.
For instance, the identifier function for square structures is defined as
\begin{equation} \label{eq:state_identifier}
	p(s, \psi_4, \psi_6) = 
	\begin{cases}
		1, \quad \mathrm{if} \quad \psi_4 > \psi_4^\thr, \psi_6 \leq \psi_6^\thr	\\
		0, \quad \mathrm{else}
	\end{cases} .
\end{equation}
\subsection{Classification of dynamical states} \label{app:transition_classification}
Under oscillatory shearing, the particle layers alternate periodically between the three structures s, d and h defined in the previous section.
We refer to these periodic alternations as dynamical states.
Depending on $\dot{\gamma}_0$ and $\omega$, a dynamical state may involve only one of these structures (s, d, h) or a mixture of two or all three of them during a half-period $\tau_\omega/2$.
Ignoring the (absent) forbidden case, we therefore allow the detection of three \textit{pure} and four \textit{mixed} dynamical states (see \tab\ref{tab:transition_classification}),
\begin{table}[t]
	\caption{Criteria for the classification of dynamical states. 
		Unmentioned structures in the last column are adopted less than $1\%$ of the time during each period. 
		Dynamical states in parenthesis are not observed.}
	\label{tab:transition_classification}
	\begin{tabular}{l|c|c}
		\multicolumn{1}{l|}{dynamical state}		&	abr.	&	criterion	\\
		\hline
		square							&	s		&	$P(\mathrm{s}) \approx 100\%$	\\
		square-disordered				&	sd		&	$P(\mathrm{s}), P(\mathrm{d}) > 1\%$	\\
		disordered						&	d		&	$P(\mathrm{d}) \approx 100\%$	\\
		disordered-hexagonal			&	dh		&	$P(\mathrm{d}), P(\mathrm{h}) > 1\%$	\\
		hexagonal						&	h		&	$P(\mathrm{h}) \approx 100\%$	\\
		(square-hexagonal)				&	(sh)	&	$P(\mathrm{s}), P(\mathrm{h}) > 1\%$	\\
		square-disordered-hexagonal		&	sdh		&	$P(\mathrm{s}), P(\mathrm{d}), P(\mathrm{h}) > 1\%$
	\end{tabular}
\end{table}
of which the dynamical state sh is not observed.
This is because the square and hexagonal sectors can only be connected by crossing the disordered sector (compare \fig\ref{fig:traj_transition_states}).
A dynamical state is considered ``mixed'' if each of the participating states is assumed at least 1\% of the time across multiple oscillation periods and different ensembles.

Mathematically speaking, we define the average ratio of time spent in state $\alpha$
\begin{subequations} \label{eq:ratio_time_spent}
\begin{align}
	P(\alpha) 
	&= \frac{1}{N} \sum_{i=1}^N \frac{1}{(T-t_\mathrm{s})} \int\limits_{t_\mathrm{s}}^{T} p(\alpha, \psi_4^i(t), \psi_6^i(t)) \mathrm{d}t ,	\\
	&= \frac{1}{N} \sum_{i=1}^N \frac{1}{N_{\Delta t} - j_\mathrm{s}} \sum_{j=j_\mathrm{s}}^{N_{\Delta t}} p(\alpha, \psi_4^i(j), \psi_6^i(j)) ,
\end{align}
\end{subequations}
where $T$ is the total simulation time, $\Delta t$ the finite time step size, $N_{\Delta t}=T/\Delta t$ the total number of time steps, $N$ the number of ensembles, $i$ the ensemble index and $j$ is the time step. 
Finally, $t_\mathrm{s}$ is the structure settling time and $j_\mathrm{s} = N_{\Delta t} t_\mathrm{s}/T$ the corresponding time step. 
Since we are lacking a proper definition of $t_\mathrm{s}$, we approximate it via $t_\mathrm{s} = \frac{2}{3} T$.
\section{System size dependence and structural heterogeneities}
\label{app:system_size}
In the following section, we comment on the system size dependence of the phenomena presented in this study based on few test simulations with a four times larger system (comprised of 4232 instead of 1058 particles).

Generally, we find that larger systems tend to exhibit more disordered structures during an oscillation cycle.
In particular, dynamical states involving hexagonal (h) structures at intermediate frequencies seem to be affected, i.e. the dynamical states sdh, dh and h.
For example, at $\dot{\gamma}_0 \tau_\mathrm{B}=769, \omega \tau_\mathrm{B}=51.5$ (see \fig\ref{fig:traj_transition_states}d), a system with 1058 particles exhibits a dh-state, assuming a disordered structure for 8\% and a hexagonal structure for 92\% of each oscillation period.
In contrast, the corresponding larger system with 4232 particles assumes a disordered structure 33\% and a hexagonal structure 67\% of each period, which remains a dh-state but with more disorder.

According to this shift, some h-states turn into dh-states and some dh-states into d-states for larger system sizes compared to smaller system sizes.
Generally speaking, the order d-dh-h (from small to large $\dot{\gamma}_0$), which we observe at intermediate frequencies (compare \fig\ref{fig:sd_dynamical_states}), remains valid, but the transition lines get shifted to higher $\dot{\gamma}_0$.
Similarly, some sd-states turn into d-states and sdh-states into either sd-, dh- or d-states, although states including square structures tend to be less affected.
On the other hand, the transition from s- to sd-states as well as the depinning transition seem to remain unaffected.

\begin{figure}
	\centering
	\includegraphics[width=\linewidth]%
		{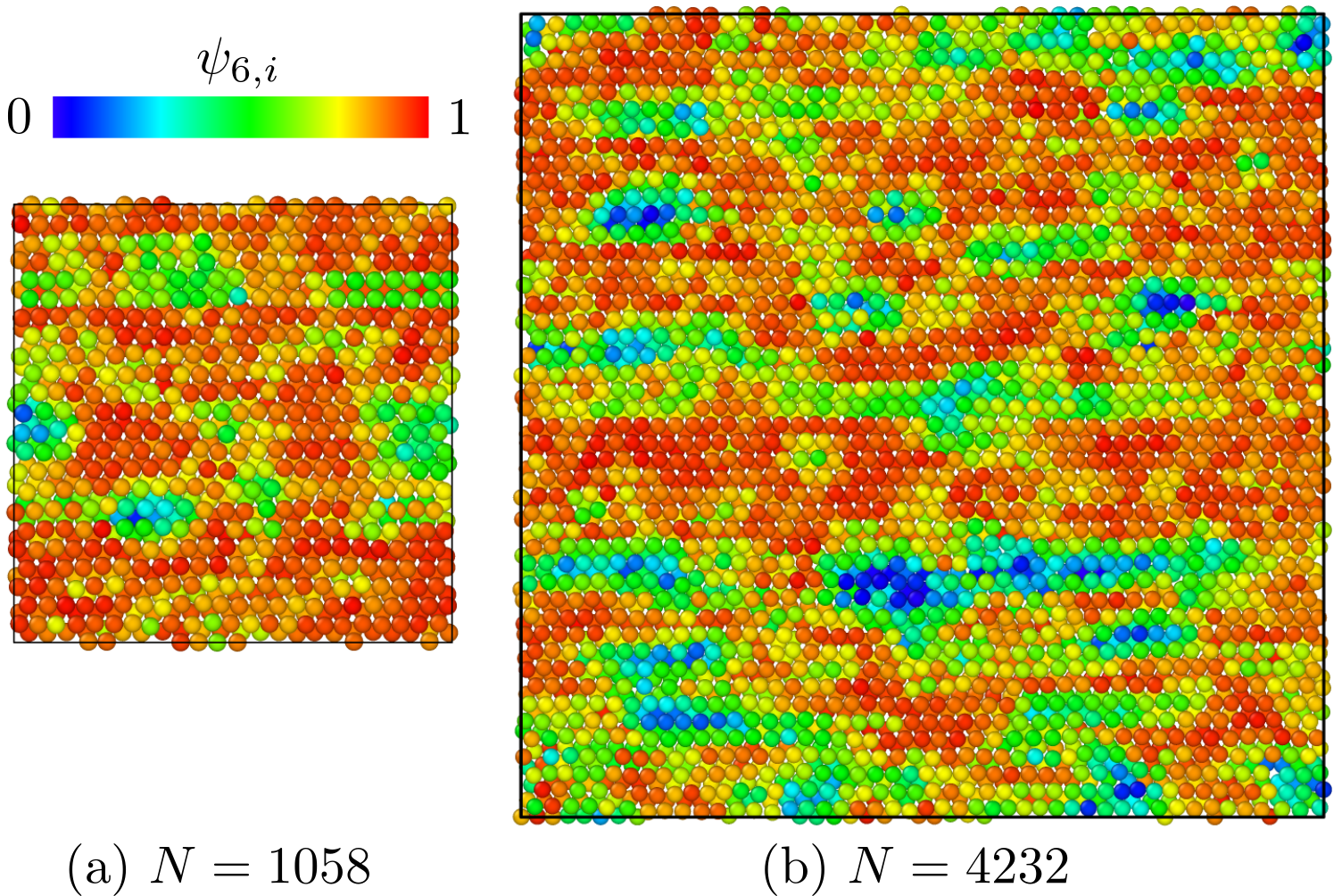}
	\caption{
		Snapshots of two disordered structures from the dh-state shown in \fig\ref{fig:angBond_example} ($\dot{\gamma}_0 \tau_\mathrm{B} = 769, \omega \tau_\mathrm{B} = 51.5$) for a smaller (a) and a larger (b) system size.
		The snapshots are taken at $t = 4.75 \tau_\B$ when the structures change from disordered to hexagonal.
		Particles are colored based on their individual $\psi_{6,i}$-value, which (partly) determines their local structure (blue=square, green=disordered, red=hexagonal).
		Although the snapshots are taken at the same point in time, the smaller system is already more hexagonal compared to the larger system, because some hexagonal stripes have reinforced themselves by connecting with their periodic images.
	}
	\label{fig:system_size}
\end{figure}

Upon a closer look at the particle-resolved angular bond order parameters, we find that the reason for a preference of d- over h-structures at larger system sizes lies in structural heterogeneities.
In fact, it turns out that (on average) disordered structures are comprised of a mixture of square and hexagonal substructures.
We find that these substructures typically have the shape of elongated stripes along the shear ($x$)-direction (compare \fig\ref{fig:system_size}), 
which for a smaller system size can quickly cover the whole width of the simulation box.
It further seems that these (hexagonal) stripes reinforce their structure once they connect with their periodic image, thereby delaying the transition back to disordered structures.
For a larger simulation box on the other hand, it is less likely that a stripe connects with its periodic image.
Hence, it takes longer for a disordered structure to become fully hexagonal, which is necessary for the formation of dh- or h-states.

The formation of elongated h-stripes also applies to a certain extent during the transition from s- to d-structures.
We find it to be less pronounced, though, because the formed, hexagonal stripes are shorter in length and thus less likely to connect with their periodic image.

Ultimately, a larger system size does not alter the qualitative behaviour shown in \fig\ref{fig:sd_dynamical_states}, but enlarges the disordered (green) region at $\dot{\gamma}_0 \tau_\mathrm{B} \approx 360, \omega \tau_\mathrm{B} \approx 100$ and shifts the disordered-hexagonal (yellow) and hexagonal (red) regions upwards.
\normalem	
%

\end{document}